\documentclass[preprint2]{aastex}

\slugcomment{Accepted for publication in the ApJ; submitted 2020-06-09, accepted 2020-07-24.}

\shorttitle{Variability in PPNs: VII. Medium-bright, O-rich}

\shortauthors{Hrivnak et al.}

\begin{document}

\title{Variability in Proto-Planetary Nebulae: VII. Light Curve Studies of Five Medium-bright, Oxygen-rich or Mixed-chemistry Post-AGB/Post-RGB Objects}

\author{Bruce J. Hrivnak\altaffilmark{1,2}, Gary Henson\altaffilmark{2,3}, Todd C. Hillwig\altaffilmark{1,2}, Wenxian Lu\altaffilmark{1}, Kristie A. Nault\altaffilmark{1,4}, and {Kevin Volk}\altaffilmark{5} }

\altaffiltext{1}{Department of Physics and Astronomy, Valparaiso University, 
Valparaiso, IN 46383, USA; bruce.hrivnak@valpo.edu, todd.hillwig@valpo.edu, kristie.nault@valpo.edu, wen.lu@valpo.edu (retired)}
\altaffiltext{2}{Southeastern Association for Research in Astronomy (SARA), USA}
\altaffiltext{3}{Department of Physics and Astronomy, East Tennessee State University, 
Johnson City, TN 37614, USA; hensong@mail.etsu.edu}
\altaffiltext{4}{Presently at the Department of Physics and Astronomy, University of Iowa, Iowa City, IA, 52242, USA; kristie-nault@uiowa.edu}
\altaffiltext{5}{Space Telescope Science Institute, 3700 San Martin Drive, Baltimore, MD, 21218, USA; volk@stsci.edu}

\begin{abstract}

We have monitored over a ten-year interval the light variations of five evolved stars with very large mid-infrared excesses.
All five objects appear to have oxygen-rich or mixed oxygen-rich and carbon-rich chemistries.
They all vary in light: four over a small range of $\sim$0.2 mag and the fifth over a larger range of $\sim$0.7 mag.  
Spectral types range from G2 to B0.
Periodic pulsations are found for the first time in the three cooler ones, IRAS 18075$-$0924 (123 days), 19207$+$2023 (96 days), and 20136$+$1309 (142 days). 
No significant periodicity is found in the hotter ones, but they appear to vary on a shorter time scale of a few days or less.
Two also show some evidence of longer-term periodic variations ($\sim$4 yrs). 
Three appear to be proto-planetary nebulae, in the post-asymptotic giant branch (post-AGB) phase of stellar evolution.  
 Their light variations are in general agreement with  the relationships between temperature, pulsation period, and pulsation amplitude found in previously studied PPNe.
 The other two, however, appear to have too low a luminosity (1000$-$1500 L$_\sun$), based on {\it Gaia} distances, to be in the post-AGB phase.  Instead, they appear to be Milky Way analogues of the recently identified class of dusty post-red giant branch stars found in the Magellanic Clouds, which likely had their evolution interrupted by interaction with a binary companion. If this is the case, then these would be among the first dusty post-RGB objects identified in the Milky Way Galaxy.\\
 \end{abstract}

\section{INTRODUCTION}

This is the second of two rather similar papers describing our photometric study of several medium-bright ({\it V}=13$-$15 mag) proto-planetary nebula (PPN, plural PPNe) candidates \citep{hri20}.  PPNe are objects in the short-lived (few thousand years) transition between asymptotic giant branch (AGB) stars and planetary nebulae (PNe).
As such, they fall into the larger classification of post-AGB objects.

PPNe have been classified as oxygen-rich (O-rich) or carbon-rich (C-rich) based on which of the two atoms is more abundant (and consequently more dominant) in the atmospheric and circumstellar chemistry.  
Oxygen is cosmically the more abundant of the two, and stars form with more oxygen than carbon.  
However, during the AGB stage, additional carbon is produced, which through the third dredge up process can be mixed to the surface.
For stars with masses greater than approximately 1.5 but less than 3$-$4 M$_\sun$, the carbon can become more abundant than oxygen and lead to a C-rich star \citep{kar14,slo16}. 
The C-rich PPNe are in several ways the more interesting, with molecular carbon (C$_2$ and C$_3$) absorption features present in their visible spectra and unidentified infrared bands (UIR) in their infrared spectra  \citep{hri08}.  In addition, the C-rich PPNe possess greatly enhanced {\it s}-process elements in their atmospheres, an evidence of nucleosynthesis and the third dredge up on the AGB \citep{vanwin03}.  
Some also possess a strong but unidentified ``21 $\mu$m'' emission feature \citep{hri09}.
The O-rich PPNe share with the C-rich ones the chemical property of a low metal abundance but not the enhanced {\it s}-process elements.
Their mid-infrared spectra display silicate emission at 9.7 and 18 $\mu$m and crystalline silicate emission features \citep{mol02}.
There are also a smaller number of ``mixed-chemistry'' objects, showing in particular both C-rich UIR features and O-rich crystalline silicate and sometimes the 9.7 $\mu$m features in their mid-infrared spectra.

PPNe are known to vary in brightness.  
In this paper, we report on the search for periodic brightness variations in the light curves of five medium-bright PPN candidates which appear to have O-rich or mixed chemistries.
This follows our earlier variability studies of four bright ({\it V} = 7-11 mag) O-rich PPNe \citep{hri15b, hri18}. 
Most of our previous variability studies have been of C-rich PPNe in the Milky Way Galaxy \citep{hri10,hri13,hri20} and in the Magellanic Clouds \citep{hri15a}.
Light curve studies of PPNe have also been carried out by Arkhipova and collaborators \citep[e.g.,][]{ark10,ark11}.
PPNe have been found to vary in brightness with periods in the range of 35 to 160 days, but with complex light curves indicating variable amplitudes and additional secondary periods. 
We begin this present study by describing the candidate objects and our photometric observations of them, and then we present the results of the period analysis of each object.
Some recent, publicly-available, all-sky photometric data for these objects were also included in the analyses. 
We conclude with an examination of their evolutionary state and a discussion of the light curve results, comparing them with previous studies.

\section{PROGRAM OBJECTS}

The five objects in this program were chosen as PPN candidates on the basis of their strong emission in the mid-infrared, as revealed in observations made with the {\it Infrared Astronomical Satellite} ({\it IRAS}).  Such emission is a strong indicator of a detached dust shell which is warmed by absorption of light from a central star.
This is a typical of a PPN.  

These program objects are of medium brightness, with {\it V} $\approx$ 13$-$15 mag. 
They are listed in Table~\ref{object_list}, with their primary identifications being their {\it IRAS} catalog numbers.  
Listed also are their identifications in some additional star catalogs, coordinates in the equatorial and galactic systems, mean {\it V} magnitudes and color indices, and spectral types. 
The association of the medium-bright star with the {\it IRAS} source was directly confirmed for two of the five by 10 $\mu$m ground-based observations.  
These were carried out at the United Kingdom Infrared Telescope for IRAS 18075$-$0924, and 20136+1309, with the observations for the latter of these published by \citet{su01}. 
For the other three, the association is based on the close positional agreement between the visible object and the mid-infrared {\it IRAS} and {\it Wide-field Infrared Survey Explorer} ({\it WISE}) sources, and as we will see, the good agreement with the spectral classification and expected light curve properties.   

\placetable{object_list} 

These five objects were initially judged to likely be O-rich objects based on the absence of molecular carbon features in their visible spectra.
However, multi-wavelength observations do not fully support this.
OH maser emission has been detected in IRAS 19207+2023 in the main line at 1612 MHz and in the 1665 MHz line but not 1667 MHz \citep{yung14}.   In IRAS 20136+1309, a weak OH maser detection has been reported, but only in the 1667 MHz line \citep{lew00}.  
No OH detections were made in IRAS 19039+1232 and 19306+1309 \citep[respectively]{lew92,lik89}, but these non-detections may be due to a lack of sensitivity.
IRAS 19306+1407 has been detected in CO emission \citep{sancon12}. 
A mid-infrared spectrum of IRAS 19306+1309 shows both UIR (C-rich) features and the 9.7 $\mu$m silicate (O-rich) feature \citep{cer09}. 
Thus, IRAS 19207+2023 and 20136+1309 appear to be O-rich objects and IRAS 19306+1407 appears to have a mixed chemistry.  
IRAS 18075$-$0924 is presumably O-rich, based on the absence of C$_2$ absorption features in the visible spectrum of this G spectral type star, in contrast to the presence of these C$_2$ features in the spectra of C-rich F$-$G PPNe \cite[e.g.,][]{hri95}.  
While C$_2$ absorption features are also absent from the hotter, mid-B star IRAS 19039+1232, we cannot make the same claim about its chemistry since C$_2$ features are also not seen in the spectrum of the C-rich B2 PPNe candidate IRAS 01005+7910 \citep{kloch02}.  
So at present, the chemistry of IRAS 19039+1232, while likely oxygen-rich, is uncertain.

{\it Hubble Space Telescope} images have been obtained for two of the program objects.
IRAS 19306+1407 displays a faint bipolar nebula with rather cylindrical lobes around the central star \citep{sah07, siod08}.
For IRAS 20136+1309, one sees only the central star, but there is a suggestion of faint nebulosity given by the slightly extended width of the image as compared with a point source \citep{su01}.

\section{PHOTOMETRIC OBSERVATIONS AND REDUCTIONS}
\label{obs}

Most of the observations were carried out at the Valparaiso University Observatory (VUO) with the 0.4 m telescope and SBIG 6303 CCD camera.  Observations began in 2008 and were carried out through 2017 or 2018, depending upon the object.
These were complemented by observations made with two telescopes operated by the Southeastern Association for Research in Astronomy (SARA). 
They are the 0.9 m at Kitt Peak National Observatory (SARA-KP) and the 0.6 m Lowell telescope at Cerro-Tololo Interamerican Observatory (SARA-CT).  
The SARA observations began in 2009 or 2010  and continued through 2016 or 2018, depending upon the object.
The two SARA telescopes were equipped with several different CCD detectors during this time \citep{keel17}.
Standard Johnson {\it V} and Cousins {\it R$_C$} filters were used with each of the telescope-detector systems.  
One object, IRAS 20136+1309, was only observed with the VUO telescope.

The CCD images were reduced using standard procedures in IRAF\footnote{IRAF is distributed by the National Optical Astronomical Observatory, operated by the Association for Universities for Research in Astronomy, Inc., under contract with the National Science Foundation.}.
These included removal of cosmic rays, subtraction of the bias, and flat fielding of the images. 
Instrumental magnitudes were derived using an aperture of $\sim$5.5$\arcsec$ radius;
this accommodated the seeing quality of the VUO site, which was typically 2.5$-$3.0$\arcsec$ FWHM, and was similar to that used in the reduction of earlier observations
made at the VUO with a different CCD. 
The statistical uncertainty in the individual observations was $\lesssim$0.010 mag.
The observations were transformed to the standard system using stars from the lists of \citet{land83,land92}.
Standard stars were observed on selected photometric nights, along with our program stars.
Linear color coefficients were determined and used in the transformation to the standard photometric system.  
Standard magnitudes of the program stars made on the best nights are listed in Table~\ref{std_ppn}.
We carried out a program of differential photometry, using three comparison stars in each field.  
The comparison stars were chosen on the basis of being relatively isolated stars of similar brightness nearby in the CCD field of view.
These comparison stars are listed in  Table~\ref{std_comp}, along with their standard magnitudes.
They are found to be constant at the level of $\pm$0.01 to $\pm$0.02 mag over this ten year observing interval. 

\placetable{std_ppn} 

\placetable{std_comp} 

When combining the standardized data from the different telescope-detector-filter systems, we found that systematic offsets existed.  
We attribute these to the redness of the program objects (see Table~\ref{object_list}), in some cases a relatively large color coefficient (0.1$-$0.2) for a detector-filter system, and the neglect of second-order terms in the color and extinction coefficients.  
More details regarding the empirical determination of the offset values and access to the standardized data are found in the Appendix.
The offsets were generally small, between $-$0.02 and $+$0.02 mag, but not always.  Estimated uncertainties in the offset values are $\pm$0.01 to $\pm$0.02 mag.  
We expect the existence of these offset corrections to have little effect on the determination of the period or amplitude of the light variations.

\begin{figure}\figurenum{1}\epsscale{2.20} 
\plotone{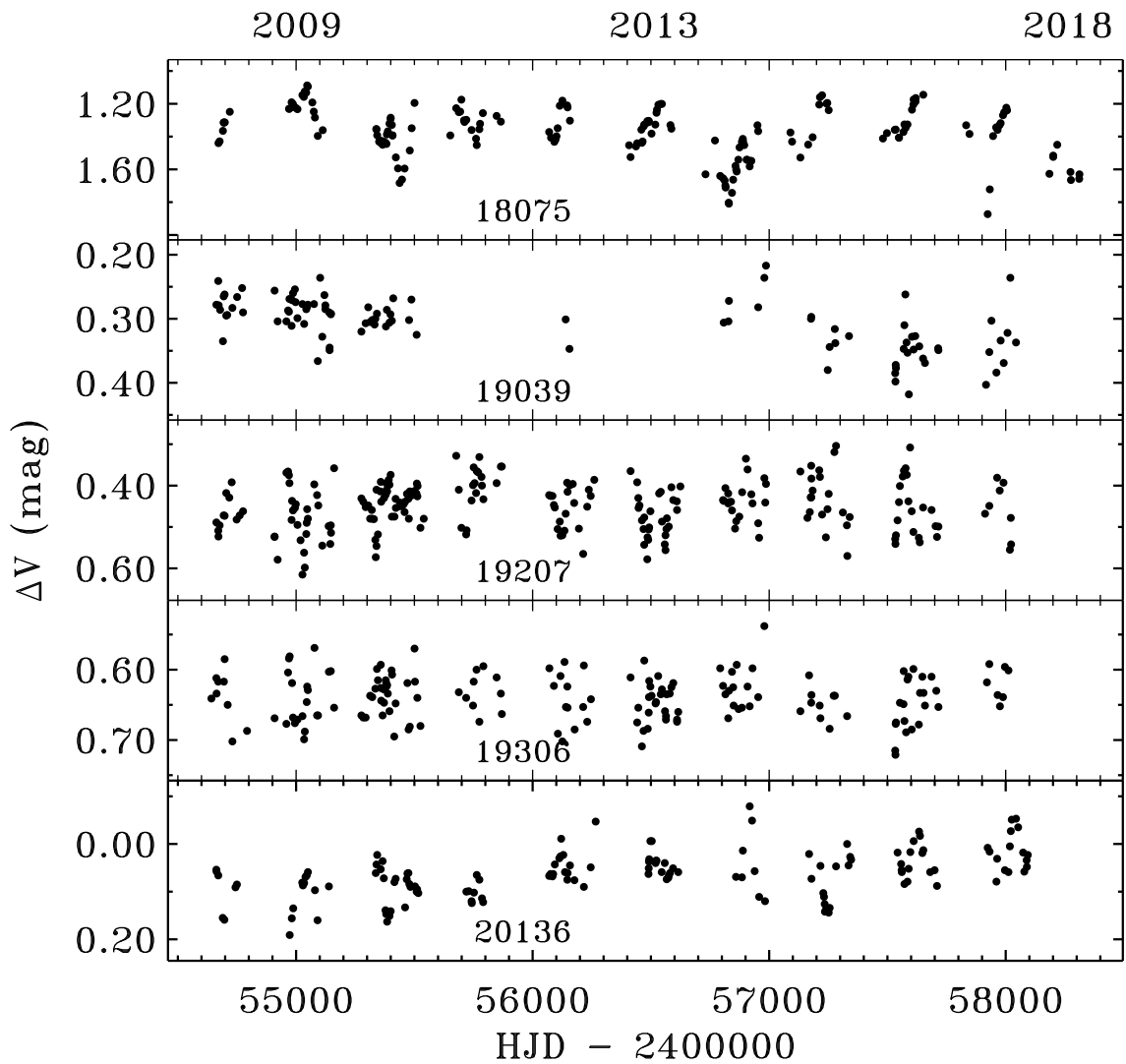}
\caption{The differential {\it V} light curves from 2008$-$2018 for the five program objects.  
Typical error bars are $\pm$0.010 to $\pm$0.020 mag and are shown in Figure~\ref{LC_fits}.
(Note that the brightness scales are not the same for all objects, with IRAS 18075$-$0924 varying over a larger range.)
\label{fig1}}
\epsscale{1.0}
\end{figure}

The standardized, differential {\it V} light curves based on the combined 2008$-$2018 data, including offsets, are shown Figure~\ref{fig1}.
Plotted are the differences in brightness between each program star and its comparison star 1 (C1).
The uncertainty in the combined data is $\pm$0.01 to $\pm$0.025 mag, with a typical uncertainty of $\pm$0.015 mag.
All of the objects vary in brightness, with variations on the order of 0.2$-$0.3 mag, except for IRAS 18075$-$0924, where the variations reach $\sim$0.7 mag over the observing interval.
The light curves of the individual objects are discussed and analyzed in Section~\ref{LC-Var}.

\placefigure{fig1}

\begin{figure}\figurenum{2}\epsscale{0.70} 
\plotone{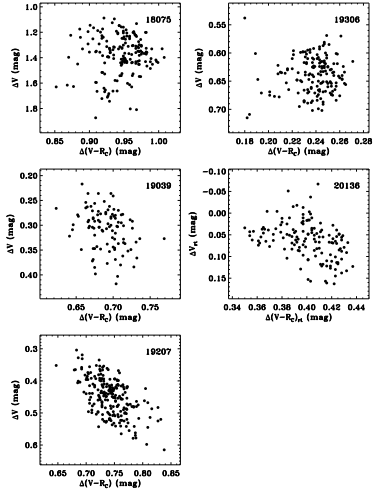}
\caption{The differential color curves for the target objects based on our new data: $\Delta$({\it V$-$R$_C$}) versus $\Delta${\it V}.  Only IRAS 19207+2023 and 20136+1309 show a clear trend of increasing color index with decreasing brightness. 
For IRAS 20136+1309, we first removed the temporal trends (rt) of increasing brightness and increasing color index from the data (see Section~\ref{20136}). 
\label{color}}
\epsscale{1.0}
\end{figure}

Since we have both {\it V} and {\it R$_C$} observations, we can examine color variations in the objects.  
In Figure~\ref{color} are shown the {\it V} versus ({\it V$-$R$_C$}) color curves based on the 2008$-$2018 data. 
The color changes are small, typically 0.10$-$0.15 mag.  Two of the objects show a general trend of being redder when fainter, 
while the others show no general trends.
These will be discussed individually.

\placefigure{color}

\begin{figure}\figurenum{3}\epsscale{2.0} 
\plotone{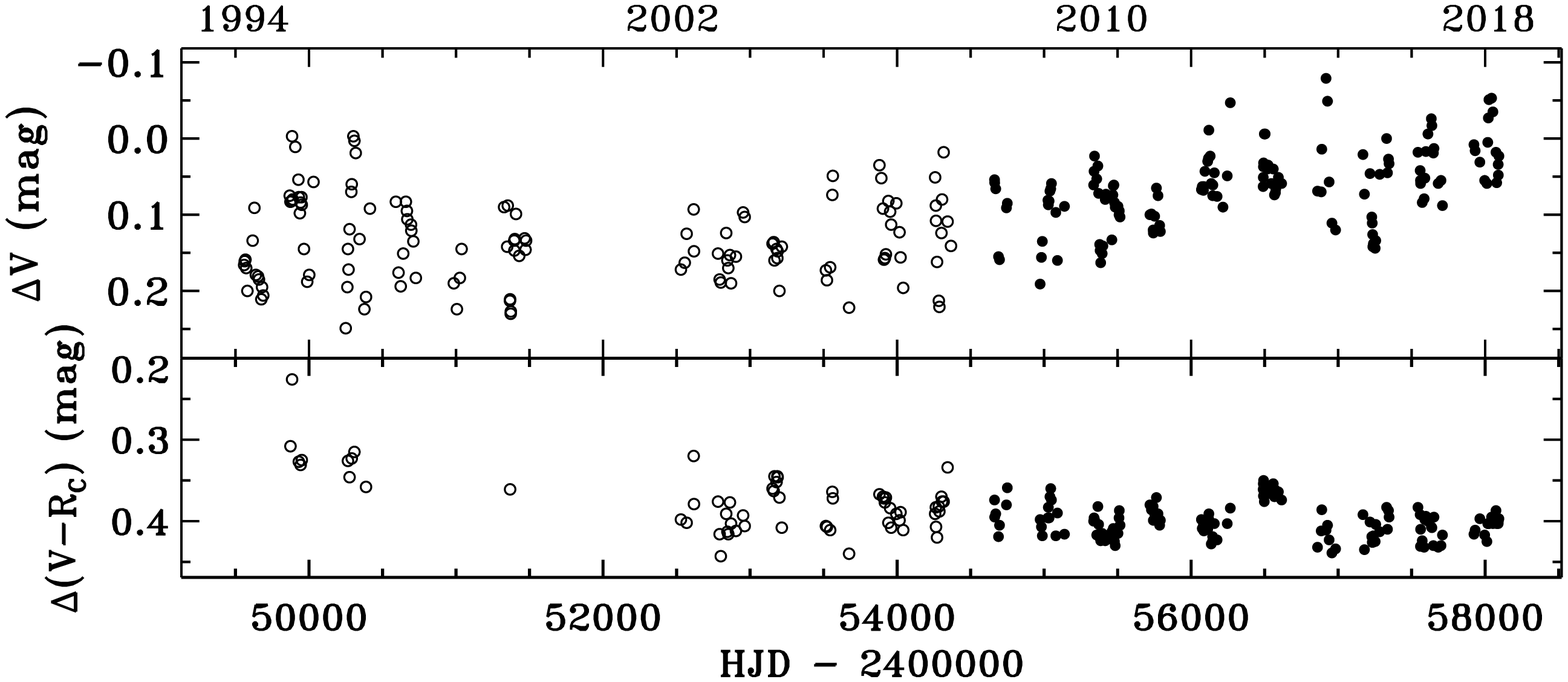}
\caption{The earlier differential {\it V} light and ({\it V$-$R}$_C$) color curves of IRAS 20136+1309 from 1994 to 2007 (open circles), combined with our more recent 2008 to 2017 data (filled circles).  These show the longer-term trends in the data. 
\label{fig2}}
\epsscale{1.0}
\end{figure}

One of the objects, IRAS 20136+1309, was also observed earlier, from 1994$-$2007, at the VUO using a Photometrics Star I CCD that did not have guiding capabilities.  That limited our integration times to $\le$6 min and resulted in somewhat less precise data, particularly in the {\it V} band.  The earlier {\it V} light curve is shown in Figure~\ref{fig2}, along with the more recent data.
The data appear to fit together smoothly without the need of an offset.  
The combined VUO {\it V} and {\it R}$_C$ light curves will be used to investigate the presence of longer-term trends in the data.  
Also shown are the associated ({\it V$-$R}$_C$) color curves from the same time interval.

\placefigure{fig2}

\section{ASAS-SN LIGHT CURVES}

\begin{figure}\figurenum{4}\epsscale{2.2} 
\plotone{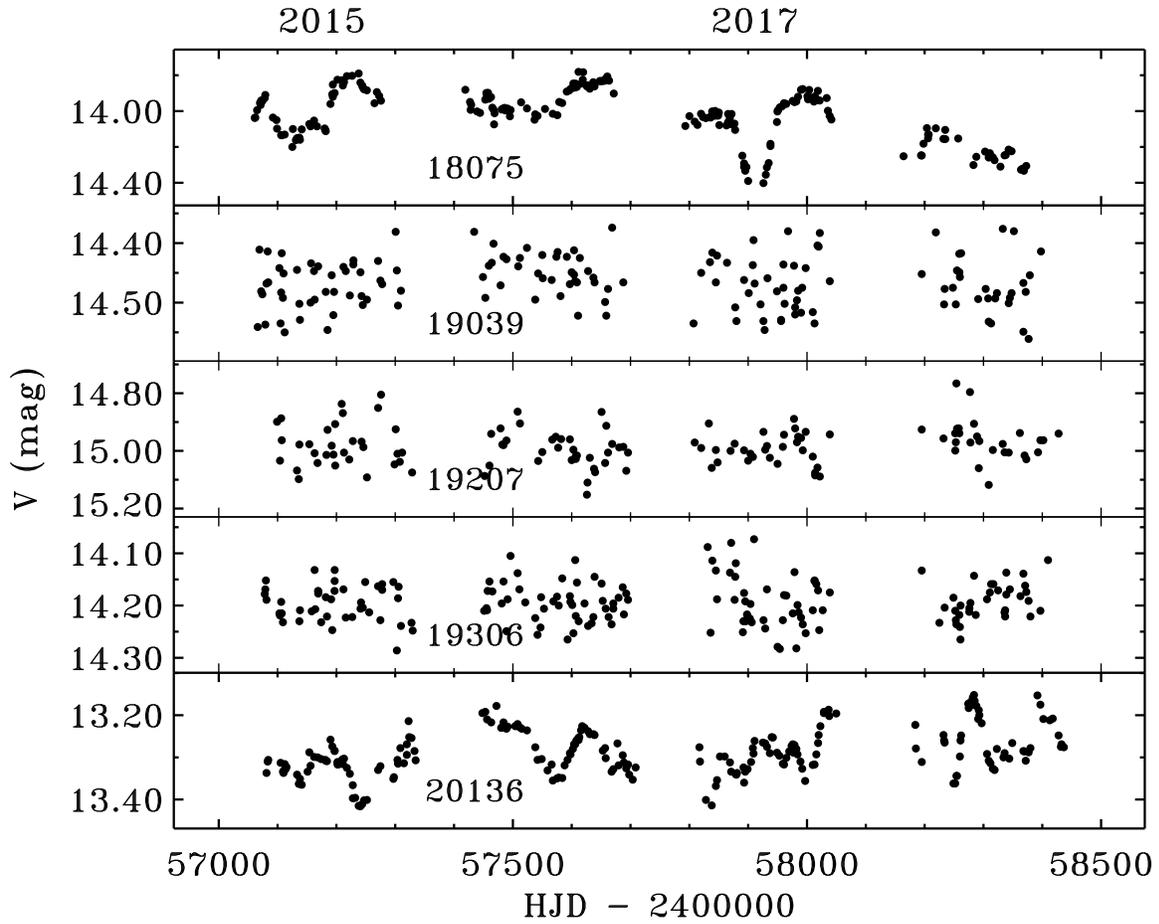}
\caption{The ASAS-SN {\it V} light curves from 2015 to 2018. 
(Note that the brightness scales are not the same for all objects, with IRAS 18075$-$0924 varying over a larger range.)
\label{fig4}}
\epsscale{1.0}
\end{figure}

Frequent observations of these objects have recently been made with the All-Sky Automated Survey for Supernovae \citep[ASAS-SN;][]{koch17}.  
This automated survey uses cameras with 14 cm telephoto lenses and CCD detectors located in good astronomical sites, and typically takes three successive dithered images of 90 s each.   
These observations effectively began in early 2015 with the {\it V} filter, and we have gathered the data on these through late 2018,
when the publicly-available {\it V} data end.  ASAS-SN {\it g} filter data (using the SDSS {\it g} filter) became available beginning in early 2018 and continue to the present, but cover a shorter interval of time.  For the two objects for which we find short timescale variability but no periodicity in the {\it V} filter data, we also analyzed these more recent {\it g} observations.
The ASAS-SN observations provide very good coverage over these shorter intervals, although the precision in the ASAS-SN data is generally lower than that of our observations. 
We first removed a few of the observations with large uncertainties and then combined the successive observations on a night into an average point.  
The average-points {\it V} light curve of each object is shown in Figure~\ref{fig4}.  
The average statistical uncertainties of theses ASAS-SN average-point observations 
range from 0.013 mag for the brightest of these, IRAS 20136+1309, which is similar to our our calculated uncertainty for this object, to 0.046 mag for the faintest, IRAS 19207+2023, which is a little more than twice as large as ours.

\placefigure{fig4}

\section{LIGHT CURVES AND VARIABILITY STUDY}
\label{LC-Var}

\subsection{PERIODOGRAM ANALYSIS}

PERIOD04 \citep{lenz05} was the primary program used to search for periodicities in the data.
It uses a Fourier analysis to search for periods in the frequency domain, and it has the advantage that it can fit more than one sine curve to the data simultaneously. 
We have used the common significance criteria of a signal-to-noise ratio (S/N) $\ge$ 4.0 \citep{bre93}.   
For the objects for which we have found periods in the data, we also used several other readily available periodogram programs to determine the primary period and found similar results. 
These procedures are described in much more detail in our complementary paper \citep{hri20}, and the reader is referred there for additional information.

In our discussion and period analysis below, we follow the order of the presentation of the light curves.  However, we make an exception and describe last the more complicated system IRAS 18075-0024.

\subsection{IRAS 19039$+$1232}

This object was observed regularly during the first three years, 2008$-$2010, then rather infrequently if at all during the next five years, then again more regularly during the last two years 2016$-$2017.  
The overall light curve shows a general decrease in brightness over these ten seasons of about 0.06 mag in {\it V} and {\it R$_C$}, and it shows a full range of brightness variations of 0.20 mag in {\it V} and 0.19 mag in {\it R$_C$}.
The peak-to-peak variability in a season varies from 0.06 to 0.16 mag in {\it V} and from 0.10 to 0.16 mag in {\it R$_C$}, confining ourselves to the five seasons with the majority of the observations.
No periodicity is evident based on visual inspection, but it can be seen that the brightness measurements change (peak-to-peak) over a short timescale of less than a week.
The ({\it V$-$R$_C$}) color varies over a range of 0.10 mag, neglecting two outliers, and shows no correlation between brightness and color (Fig.~\ref{color}).

A period analysis was carried out on the light curves, correcting first for the general decreasing trend in brightness.
No significant periods were found in the {\it V} or {\it R$_C$} data, although values of 63 and 27 days (S/N$\approx$3.4) were found in each of them.
ASAS-SN {\it V} light curve from 2015$-$2018 was also analyzed.
Over that time interval, there is no general trend in brightness.  
A period analysis determined a most likely value of 62 days, but it was not significant ({\it S/N}=3.0).
Thus, although there are suggestions of a period of around 63 days and perhaps another one about 27 days in these three data sets, these periods are not close to meeting our significance criterion (S/N$\ge$ 4.0).  
No period was found in the more recent and fainter ASAS-SN {\it g} observations.
Visual inspection suggested much shorter intervals to the variations ($<$ 1 week).  
Thus we do not claim to have found periodicity in any of these light curves of IRAS 19039+1232.

\subsection{IRAS 19207+2023}

This object varies in brightness within a season, and variations are also seen in the average brightness between seasons (Fig. 1).  
The overall brightness varies over ranges of 0.31 ({\it V}) and 0.20 ({\it R$_C$}) mag, 
with seasonal ranges of 0.16 to 0.26 mag in {\it V} and 0.13 to 0.19 mag in {\it R$_C$} and an average seasonal amplitude ratio ({\it R$_C$}/{\it V}) of 0.84. 
The average brightness between seasons varies over a range of 0.07 mag in {\it V} and 0.025 mag in {\it R$_C$}.
 (These seasonal statistics are all based on the eight seasons with 19 or more individual data points.)
Visual inspection of the light curves shows a cyclical variation with a period of $\sim$100 days in most of the seasons.
The object varies over a color range of $\Delta$({\it V$-$R}$_C$) = 0.16 mag during the 10 years of observations and appears to gradually get bluer by about 0.04 mag between years 2 and 8.   It also shows a general trend of being redder when fainter in {\it V}.

\begin{figure}\figurenum{5}\epsscale{1.0} 
\plotone{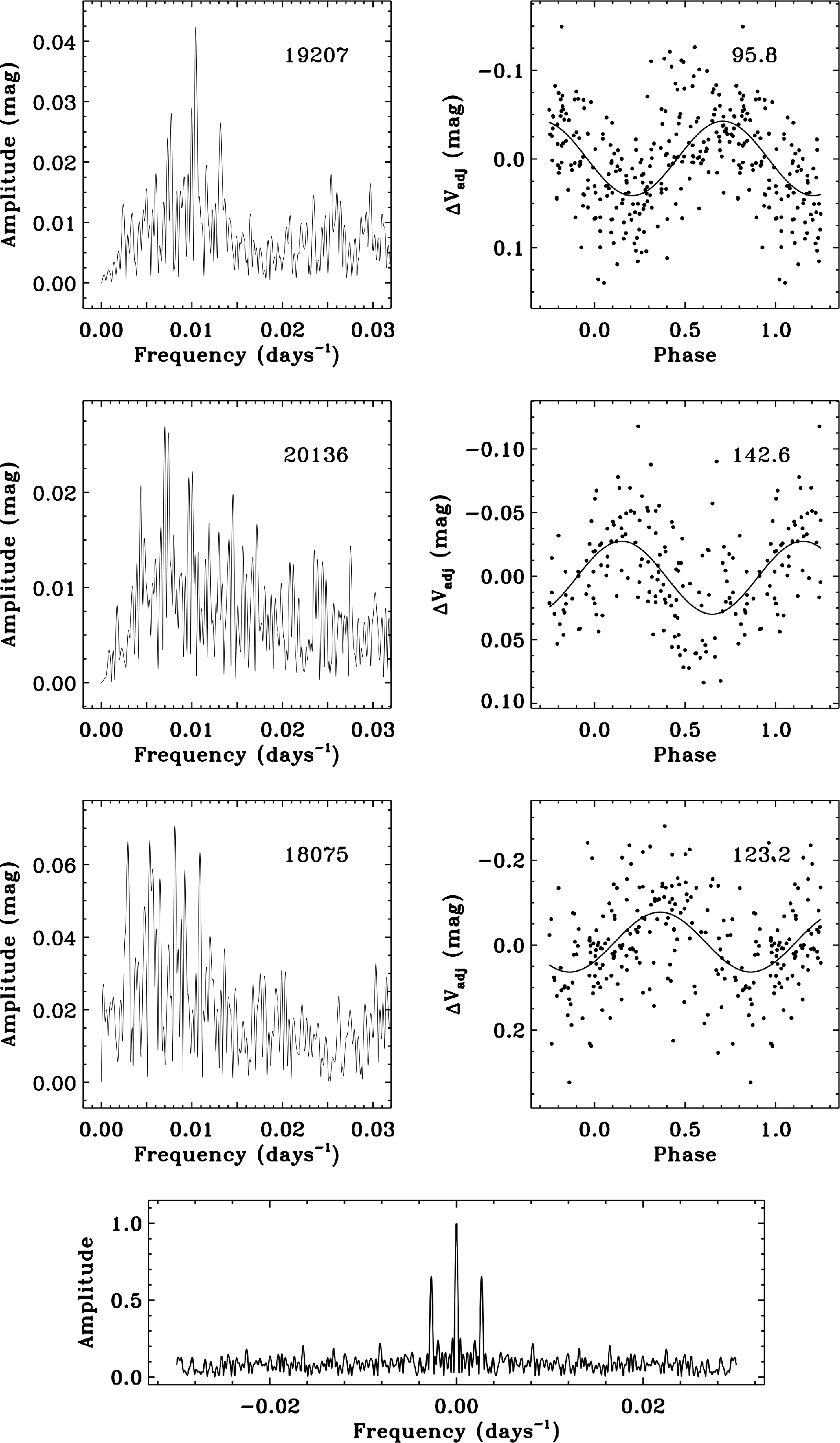}
\caption{ The frequency spectrum and the seasonally normalized {\it V} observations, phased with their dominant periods, for the three periodic objects.  The periods are listed in the phase plots on the right.  The bottom panel shows a sample frequency spectrum of the observing window for the objects.  The side lobes display the one-year alias in the data.
\label{freq_phase}}
\epsscale{1.0}
\end{figure}

\begin{figure}\figurenum{6}\epsscale{2.2} 
\plotone{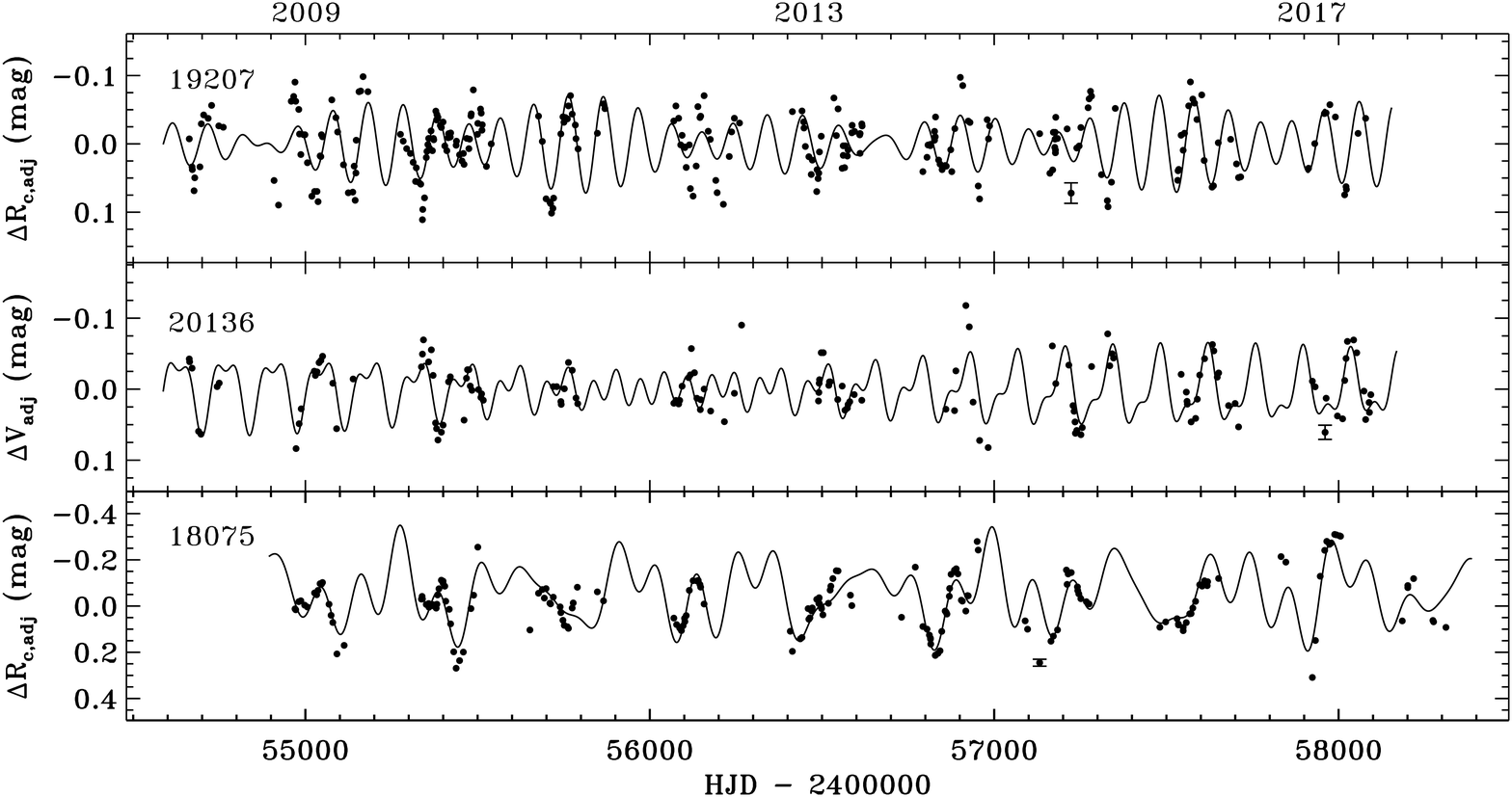}
\caption{The {\it V} or {\it R}$_C$ seasonally normalized light curves fitted by the periods, amplitudes, and phases recorded in Table~\ref{periods}.  Also shown for each is an example of a typical error bar.
(top) IRAS 19207$+$2023 from 2008 to 2017, fitted with first three periods and with a typical error of $\pm$0.020 mag; 
(middle) IRAS 20136+1309 from 2008 to 2017, fitted by three periods and with a typical error of $\pm$0.010 mag; 
(bottom) IRAS 18075$-$0924 from 2009 to 2018, fitted with four periods and with a typical error of $\pm$0.015 mag. 
\label{LC_fits}}
\epsscale{1.0}
\end{figure}

Analyses of the observed {\it V} and {\it R$_C$} light curves yield the same dominant period in each, 95.7$\pm$0.2 days (S/N=6.8).   However, in the {\it V} light curve, the second strongest period is a 4.2 year periodicity, which is a fitting to the variations in the average brightness of the different seasons.  This effect is much less in the {\it R$_C$} light curve and does not result in a significant period.  
To focus on the pulsation in this object, we removed the seasonal variations by normalizing to the average value of each season.  This resulted in the statisticly same dominant period values in each, averaging {\it P}$_1$ = 95.8$\pm$0.2 days. 
The frequency spectrum and the phased light curve for the {\it V} filter are shown in Figure~\ref{freq_phase}.
Also shown at the bottom of the figure is a sample frequency spectrum of the window function of the observations, with the side lobes corresponding to the one-year alias in the data.  
One can see the strong peak in the frequency spectrum of IRAS 19207+2023, along with the side lobes, and the good fit to the light curve phased with this period.
Additional significant periods of 100.3, 114.4, and 85.0 days are found in the {\it R}$_C$ light curve and an almost significant period of 93.2 days (S/N=3.9) is found in the  {\it V} light curve.  
These are listed in Table~\ref{periods}.
Due to the faintness and redness of this object, we have chosen to show in Figure~\ref{LC_fits} the more precise {\it R}$_C$ light curve, which is fitted well by the first three of these periods.  
Analysis of the normalized ({\it V$-$R$_C$}) color curve yielded a period of 95.9$\pm$0.4 days, the same as the dominant period in the {\it V} and {\it R$_C$} light curves, although it was not statistically significant (S/N=3.1).

\placefigure{freq_phase}

\placetable{periods}

\placefigure{LC_fits}

Regarding the long period of $\sim$4.2 yr seen in the combined {\it V} light curve (234 observations): it is also seen in the individual VUO-CCD {\it V} data (166 observations), although at a S/N = 3.8,  slightly below the significance criteria of 4.0.  Thus it is not an artifact of the combination of the different data sets with their different offsets.  
This long period is also present in the {\it R$_C$} light curve, but at a level much below significance.

The ASAS-SN {\it V} 2015$-$2018 data set for IRAS 19207+2023 was also investigated for periodicity.  The observing interval was Mar 2015 through Nov 2018.  
This object is faint ({\it V}=15) and the uncertainty in the data is correspondingly large ($\sigma$$_{\rm ave}$=0.046 mag).  Visual inspection showed a cyclical variability on the order of 100 days in only one of the four years.  However, application of PERIOD04 resulted in a marginally-significant (S/N=4.0) period of 97.6$\pm$0.7 days.  This is in fair agreement with that found in our much more precise and longer-term data set. 
This ASAS-SN data set is not long enough nor precise enough to test for the 4.2 year variability.

\subsection{IRAS 19306+1407}

This object clearly varies, with a full range of brightness variations of 0.14 in {\it V} and 0.16 mag in {\it R$_C$} (neglecting a few outliers), and with typical seasonal ranges of $\sim$0.10 mag in {\it V} and 0.12 mag in {\it R$_C$}.   
There is not much change in the average seasonal values in the {\it V} and {\it R$_C$} light curves ($\le$0.03 mag).  
No periodicity is obvious by visual inspection of this light curve.
The ({\it V$-$R$_C$}) color varies over a range of  0.09 mag during our ten years of observations.
However, most of the color data are found within a range of only 0.05 mag.
No correlation is found between brightness and color.
 
A period analysis of the observed {\it V} and {\it R$_C$} light curves each resulted in a period of 25.4$\pm$ 0.1 days, but at a S/N value of $\sim$3.5, which is well below the level of significance in each case. 
Analyzing the seasonally-normalized gave a similar result. 
We also performed a period analysis of the higher density ASAS-SN {\it V} light curve data from 2015$-$2018 ($\sigma$$\le$0.040, 173 points).  This yielded a period of 44.7$\pm$0.2 days, with a S/N value (3.9) just below significance.   However, this gave a poor fit to the light curve.  
A re-analysis using the most precise of the ASAS-SN {\it V} data ($\sigma$$\le$0.025, 139 points) yielded a much less significant result (S/N=3.4) for the same period.  Thus we do not regard it as a real physical period of the data. 
No period was found in the more recent {\it g} observations.
Testing the period of 25.4 days found from our data with the ASAS-SN {\it V} data also gave a very poor fit.
Visual inspection of the higher-density ASAS-SN light curve reveals evidence that the light varied on a much shorter timescale, varying from one extrema to the other over intervals of only 2 to 4 days.  
The above periods are both formally not significant and visually appear to be too long.  Thus we do not claim to have found a period in the ASAS-SN data nor in our data.

\subsection{IRAS 20136+1309}
\label{20136}
 
Our observations of this object are exclusively from VUO, beginning in 1994.  The earlier, 1994$-$2007 data clearly show variability, reaching a maximum seasonal range of 0.25 mag in {\it V} and somewhat less in {\it R$_C$} (Fig.~\ref{fig2}).  The data after 2001 show a general increase in brightness, which has continued with the newer data.
These older data are somewhat less precise, but still of good quality.  We estimate the uncertainties in the standardized, differential magnitudes to be $\pm$0.010 mag for the 2008$-$2017 data and $\pm$0.015 mag for the 1994$-$2007 data. 

An examination of the newer 2008$-$2017 data shows a general increase in brightness of approximately 0.11 mag in {\it V} and slightly more in {\it R$_C$}, with the object becoming slightly redder ($\sim$0.02 mag), over that ten year interval.
The object shows cyclical variability in a season, with an average peak-to-peak range in {\it V} of 0.12 mag and a maximum of 0.20 mag in 2014; the range in {\it R} is about 0.9 times that in {\it V}. 
Despite the small observed reddening of the object as it gradually increased in brightness with time, the dominant trend in the individual observations is for them to get redder when fainter, as seen in Figure~\ref{color}.
The less precise earlier observations suggest that the temporal color trend extends back to the early 1994$-$1996 observations, with the object being even bluer then.

Prior to conducting the period search of the 2008$-$2017 light curve, a linear trend was first removed from the data and then the data were normalized seasonally to bring the average values to the same level for each season.  
The {\it V} light curve reveals a dominant period of 142.3$\pm$0.4 days (S/N=4.8), with additional strong periods of 135.7 and 68.8 days and two additional significant periods of 125.7 and 71.0 days.  
The {\it R$_C$} light curve has a dominant period of 135.1$\pm$0.4 days (S/N=4.5), with additional periods of 142.8 and 70.1 days and then two additional significant periods of 58.1 and 125.8 days.  The results based on the three stronger peaks are listed in Table~\ref{periods}.
Thus the periods are similar in the two filters, but with the first and third reversed, except for shortest period in the {\it R$_C$} light curve.  The frequency spectrum and the phase plot based upon the dominant period of the {\it V} light curve is shown in Figure~\ref{freq_phase}.  
 One can clearly see in the frequency spectrum the two closely spaced and similar high peaks at 142 and 135 days, with their side lobes, and the additional peak at 69 days ({\it f}=0.0145 days$^{-1}$); the frequency spectrum for the {\it R$_C$} light curve looks similar, but with the heights of the two high, close peaks reversed.
Upon examining the first three periods in each of the light curves, one sees that each light curve has a dominant period and a similar third period that differs by 5\% and then another one that is about half the value of the two longer ones.  This combination has the effect of modulating the amplitude of the light curve (the two similar periods) and also adding a smaller amplitude secondary minimum between the deeper minima.   
The {\it V} and {\it R$_C$} periods of 142 and 135 days have similar phases in the two light curves. 
The three periods give a reasonably good fit to the {\it V} and {\it R$_C$} light curves, with the fit to the larger amplitude {\it V} light curve shown in Figure~\ref{LC_fits}.
The ({\it V$-$R$_C$}) data reveal a barely significant (S/N=4.1), longer period of 254 days, which is not seen in the {\it V} or the {\it R$_C$} light curves.

We next analyzed the older VUO light curves from 1994$-$2007.
We focused on the {\it V} light curve since it has more data. 
Because the light curve shows an increase in brightness after 2004, the data were first seasonally normalized and then we retained only the seasons with ten or more observations.  This resulted in eight seasons.  Two significant periods were found in the data, 
126.7$\pm$0.02 and 138.0$\pm$0.04 days.  
These periods are similar to the two longer periods found in the 2008$-$2017 light curves, but average 5\% shorter. 
We then analyzed together the two VUO {\it V} data sets, giving all the individual observations equal weight.
The data were seasonally normalized and we again excluded the seasons with fewer than ten data points.  This resulted in our using data from 17 seasons from 1994 through 2017.  The analysis resulted in three significant periods in the data, 143.2$\pm$0.2, 127.0$\pm$0.1, and 135.1$\pm$0.2 days.  
The longer of these is the dominant period in the 2008$-$2017 interval and the shorter one is the dominant period in the 1994$-$2007 interval, and a value similar to the middle one is seen in each of these two time intervals. 
This suggests that the dominant period may have increased over this time period.

We also examined the ASAS-SN {\it V} light curve for IRAS 20136+1309 with its high density of observations with relatively good precision ($\sigma$$\le$0.025 mag).  
Patterned variations in the light curve can easily be seen visually.
An analysis of the 2015$-$2017 ( 171 observations over three seasons), seasonally normalized, revealed a dominant period of 146.3$\pm$1.1 days (S/N=5.4), similar to but slightly longer than that found in the VUO data over the ten-year (2008$-$2017) interval.  There is also a second period of 320 days, about twice that of the first period; this has the effect of modulating the depths of the minima, so that they are approximately alternately deeper or shallower.  
For comparison, the VUO {\it V} light curve over those same three years, with many fewer observations (47), yielded a similar dominant period of 143.5$\pm$1.2 days (S/N=5.9) and a secondary period of 69.2 days.
However, the ASAS-SN {\it V} light curve for 2018 differs from the other years in that it appears to show several more closely spaced minima.  If it is included in the analysis, it leads to a dominant period of 59 days (S/N=4.0; this was also seen as a weaker period in the VUO {\it R$_C$} light curve).  There are additional periods of 136 and 156 days (which average to the dominant period of 146 days seen in the ASAS-SN light curve from 2015$-$2017).  

These light curves for IRAS 20136+1309 are complex, and our longer time scales of 10  and 24 years are helpful in determining the basic periods of the star.  They also provide some evidence that the dominant period may have increased over that time interval, but the complexity of the light curves makes that uncertain in our judgement.

\subsection{IRAS 18075$-$0924}
 
This object was observed over 11 seasons, 2008 to 2018.  
It displays relatively large variations in brightness, both among the individual observations within a season and in its average brightness level from season to season (Fig.~\ref{fig1}).
Seasonal variations range from 0.25 to 0.65 mag in {\it V} and from 0.21 to 0.62 mag in {\it R}$_C$ (for seasons with 10 or more observations).  
The variations in {\it V} are usually larger than in {\it R}$_C$, with an average seasonal amplitude ratio ({\it R$_C$}/{\it V}) of 0.91.
Three of the seasons, 2010, 2014, and 2017, show much larger decreases in brightness than the other eight seasons.
The peak-to-peak range  over the 11 years is 0.78 mag in {\it}V and 0.80 mag in {\it R$_C$}.
Visual inspection of the data clearly shows cyclical variations with a period of $\sim$80$-$120 day.  
The light curves in 2009, 2010, 2013, and 2014 
display two minima with alternating deeper and shallower depths, reminiscent of an RV Tauri-type light curve, but with the depth of the deeper minima varying by more than a factor of two among the four seasons and no pattern of every other one being the deeper.. 
The object varies over a color range of $\Delta$(V$-$R$_C$) = 0.16 mag, with most of the data varying within a range of 0.10 mag.  
It is worth noting that it is bluer by $\sim$0.04 mag in 2017$-$2018, when the overall brightness is fainter.  
However, as shown in Figure~\ref{color}, there is no simple overall correlation of brightness and color for the object.

Analysis of the observed light curves with PERIOD04 revealed the strongest period to be 3.54$\pm$0.07 yr (S/N=6), which attempted to fit the timing of the three deep minima, with another period of 2.06$\pm$0.05 yr, which helped to fit the overall variations in the average seasonal brightness levels.  Additional significant periods fitted the shorter-term variations seen in the data.  
Whether some of these longer-term variations in the seasonal brightness levels arise from the star itself or from changes in the dust obscuration is unclear.
To further study the pulsational variations, we normalized the data to the seasonal averages before we carried out the period analysis.  This could be done with reasonable confidence except for the observations of the first year (2008), in which the data do not pass through a maximum and minimum value, and of the 2017 season. 
For the 2017 observations, which possessed the largest range but with only two of the points in the faint half of the range, we manually normalized the light curve to the mean value of the two extrema.
A periodogram analysis was then carried out of the seasonally normalized observations, neglecting the first year.  
Four significant periods were found in each of the {\it V} and {\it R$_C$} data.  The period values in {\it V} and {\it R$_C$} are the same within the uncertainties: {\it P}$_1$ = 123.0$\pm$0.2, {\it P}$_2$ = 346.6$\pm$2.0, {\it P}$_3$ = 107.7$\pm$0.2, and {\it P}$_4$ = 208.9$\pm$1.1 days. 
In Figure~\ref{freq_phase} are displayed the frequency spectrum for the {\it V} light curve, along with the light curve phased to {\it P}$_1$. 
There is no dominant period seen, but the periods {\it P}$_1$ (frequency {\it f}=0.0081 days$^{-1}$) with its side lobes,  {\it P}$_2$ ({\it f}=0.0029 days$^{-1}$), and  {\it P}$_3$ ({\it f}=0.0093 days$^{-1}$) are evident in the frequency spectrum. 
The phase plot using the strongest period of 123 days displays reasonable evidence of this period in the data, but it is clear that it alone does not provide a good fit to this complicated light curve.
These four periods are listed, with their amplitudes and phases in Table~\ref{periods}, along with the standard deviations of the observations from the multi-periodic sine curve fit.
These result in a reasonably good representation of the seasonally normalized light curve and its times of maximum and minimum.
This is shown in Figure~\ref{LC_fits}, where we have fitted the more precise {\it R}$_C$ observations of this red object.
The period of 123 days fits the width of the minima seen in the light curves and the period of 346 days serves to enhance the strength of the deeper of the yearly minima.  

The ASAS-SN {\it V} light curve documents the variations over the latter four seasons (2015$-$2018) in more detail (Fig.~\ref{fig4}).  It shows clearly the decrease in brightness over the last two years.
However there are not enough seasons to get a reliable period from these data.
We also investigated the early ASAS-3 {\it V} light curve from 2001$-$2009, using only the best quality of this data set (grade=A, $\sigma$ $\le$0.05 mag, $\sigma$$_{\rm ave}$=0.04 mag, 187 data points).  Analysis of the data found a significant period (S/N=4.6) of 237$\pm$2 days, although it did not appear to be a good fit to the observations.  There was a suggestion of a longer period in the data of 2.4 yr, but it was not significant (S/N = 3.8).

Thus we find that the light curves of IRAS 18075$-$0924 are particularly complicated and do not reveal a strong, dominant period as seen in IRAS 19207+2023 or a pair of strong periods as seen in IRAS 20136+1309. 
 Instead, the strongest was a longer period of 3.5 years and then possibly one of 2.1 years.  These helped to fit the overall trends of the light curves but not the cyclical variations seen within a season.
The results based on our seasonally normalized 2009$-$2018 light curves appear to give reasonably good fits to the observations, with the strongest period of 123 days likely arising from a similar mechanism as those operating in the other two stars listed above.  A larger number of seasons with good quality data, as we have for IRAS 20136+1309, would presumably be helpful in sorting out the robustness of the multi-year period(s) and perhaps strengthen the case for the shorter pulsation periods found in IRAS 18075$-$0924.  

\section{DISCUSSION}

\subsection{Spectral Energy Distributions}

\begin{figure}\figurenum{7}\epsscale{1.0} 
\plotone{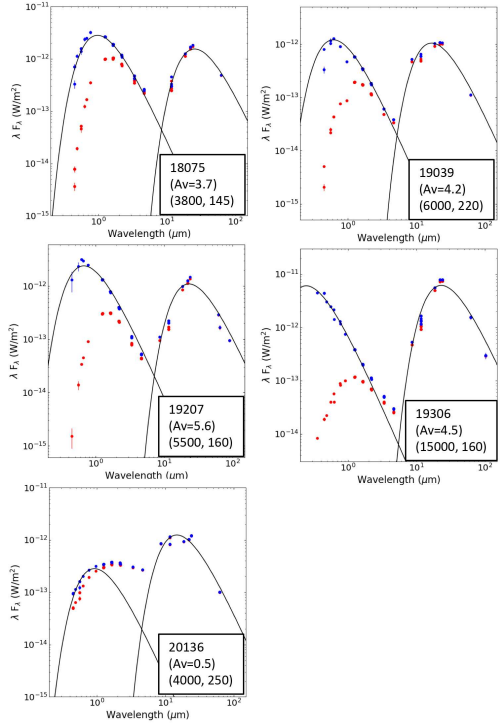}
\caption{The spectral energy distributions for each of the program objects. 
In red is plotted the observed flux density and in blue the values de-reddened for interstellar extinction.  The visual interstellar extinction is listed for each, using the upper limit.
Also drawn are two black body curves, one representing a fit to the photospheric emission de-reddened for interstellar (but not circumstellar) extinction. and the other the fit to the circumstellar dust emission. The temperatures of these are listed in the panels. 
Plotted are the {\it IRAS}, {\it AKARI}, {\it WISE}, {\it 2MASS}, Johnson {\it B,V,J,H,K}, Sloan Digital Sky Survey, and {\it IPHAS} {\it r,i} \citep{bar14} data, with error bars, as listed in SIMBAD-VizieR.
\label{SEDs}}
\epsscale{1.0}
\end{figure}

All five of the program objects show the bimodal spectral energy distributions (SEDs) characteristic of PPNe.  
They each have a peak in the visible region due to the photosphere of the star, but with the peak closer to 1 $\mu$m due to circumstellar and interstellar reddening.  The second peak occurs in the 20$-$25 $\mu$m region and is due to re-radiation from the cool dust.     
These are shown in Figure~\ref{SEDs}.   
These mid-infrared peaks vary in prominence from slightly higher than the visible peak (IRAS 18075$-$0924) to a factor of $\sim$60 times higher (IRAS 19306+1407).  
The differences are presumably due to the combination of the particular interstellar extinction and circumstellar extinction of each.

\placefigure{SEDs}

To further investigate the evolutionary state of these objects, we can use these SEDs, together with their distances, to determine approximate luminosities of the objects.
All five have distance determinations from the {\it Gaia} satellite \citep[{\it Gaia} Data Release (DR2);][]{bai18}, albeit with relatively large uncertainties (30$-$50$\%$).  
Their distances are in the range of 3 to 6 kpc.
The flux density received at the Earth can be calculated by integrating their SEDs.
However, corrections need to be made for the interstellar extinction.  These are substantial, since four of the objects lie near the plane of the Milky Way Galaxy (MWG), with $ \left|{\it b}\right| $ $\leq$ 5$\arcdeg$..
Values of the extinction in the {\it V} bandpass, {\it A}$_V$, were derived from several studies.
The study of \citet{cap17}, based on {\it Gaia} Data Release 1 distances and the combination of several datasets, produced 3D reddening maps.  However, reddening in the direction of our objects was only determined to distances of 2.0 to 3.0 kpc for the four objects near the galactic plane and 1.4 kpc for the fifth object.  These are much less than their {\it Gaia} DR2 distances, and thus yielded only lower limits to the reddening.
Upper limits for the interstellar {\it A}$_V$ in the direction of the star were determined from the the studies of \citet{sfd98} and \citet{sf11}, since these were based on observations of background galaxies.
The observed SED values for each star were de-reddened using both of these extinction values and assuming a reddening value ({\it A}$_V$/{\it E}({\it B$-$V})) of 3.1.
We applied the reddening function of \citet{rie85}, which has the virtue of extending to 13 $\mu$m and including the effect of the 10 $\mu$m silicate feature; one of us (KV) extended this function to include longer and shorter wavelengths.  
In Figure~\ref{SEDs} are also shown the observations de-reddened using the the upper limit interstellar extinction values from \citet{sf11}.
Also included in this figure are black body curves to represent the the de-reddened photospheric and the circumstellar dust emission.  These are representative curves fitted by eye.  The black body temperatures range from 3800 to 15,000 K for the photospheres, but in each case they are significantly less than that expected based on the spectral type of the star (see Table~\ref{object_list}), even though we used the upper limit to the visible extinction values.  This attests to the presence of significant circumstellar extinction in each case. 
However, we need not correct for this circumstellar dust extinction when calculating the luminosity, since it is re-radiated back into our line of site in the mid-infrared.
The dust temperatures range from 145 to 250 K.

The calculated luminosities are listed in Table~\ref{sed_table}.  
The implicit assumption in this is that the circumstellar envelopes of the objects are spherically symmetric.  If this is not the case, then particularly the visible light will be somewhat over- or under-estimated.  
We expect in these objects that the dust is optically thin in the mid-infrared; hence any asymmetries will not affect the infrared flux significantly.
Assuming that the luminosities of the objects lie approximately midway between the two interstellar extinction estimates, one finds that the luminosities range from approximately 1000 to 11,000 L$_{\sun}$.
The largest uncertainty in these values derives not from the uncertainty in the extinction or the geometry but rather in the distance.
The uncertainties in the distances for the objects have a range of about 30$\%$ lower to about 50$\%$ higher than the determined distance.  This propagates to an uncertain in the luminosity of approximately 0.5 to 2.3 times the determined luminosity.
These uncertainties are expected to diminish in future {\it Gaia} data releases.

\placetable{sed_table}

Two of the program objects have luminosities in the range of 1000 to 1500 L$_{\sun}$ based on {\it Gaia} (DR2) distances, which appear rather low for PPNe.  The other three have values in the range of 4000 to 11,000  L$_{\sun}$, values more in agreement with what one would expect based on theoretical models for PPNe \citep{milb16,blo95,vas94} and from the luminosities of the central stars of PNe \citep{gon20}.
To investigate the question of the luminosities of PPNe further, we calculated in a similar manner the luminosities of the three C-rich objects in our recent companion study \citep{hri20} and the four O-rich objects in an earlier study \citep{hri15b}.  The results for these additional objects are included in Table~\ref{sed_table}.  They cover a large range, from approximately 3000 to 28,000  L$_{\sun}$.
In discussing these luminosities, one must bear in mind the range of uncertainties cited above.

New stellar models for the evolution of stars through the post-AGB phase have been published by \citet{milb16}.
They result in post-AGB luminosities of 2500 to 25,000  L$_{\sun}$ for core masses of 0.53 to 0.83 M$_{\sun}$, respectively.  
Based on observational studies of the dusty (infrared-excess) post-AGB candidates in the SMC \citep{kam14} and LMC \citep{kam15}, \citet{kam16} have drawn attention to a new class of low-luminosity (100$-$2500  L$_{\sun}$) dusty post-red giant branch (post-RGB) stars.  Their sample was from the Magellanic Clouds, where the distances are well known.  
\citet{kam16} make a persuasive case that these objects evolved in binary systems in which the RGB star's evolution was interrupted by interaction with its companion.  This led to Roche lobe overflow or a common envelope evolution, with the star subsequently evolving toward higher temperatures, analogous to the post-AGB phase, but at a lower luminosity.

The two low-luminosity stars in this study, IRAS 19039+1232 and 20136+1309, appear to fit well into this category of dusty post-RGB stars, rather than post-AGB PPNe.  
To be post-AGB objects, they would need to be at nearly the extreme upper limits of the uncertainty of their distances, which would result in luminosities of 2400 and 3300 L$_{\sun}$, respectively, and even then the luminosity of IRAS 19039+1232 would be somewhat low.  Thus we think if far more likely that they are post-RGB objects.
The other three program objects in this study fit well into the range expected for PPNe, as do most of the other eight comparison objects in Table~\ref{sed_table}.  
IRAS 23304+6147 (L$\approx$28,000  L$_{\sun}$) and to a lesser extent IRAS 17436+5003 (L$\approx$20,000  L$_{\sun}$) stand out as being highly luminous for post-AGB objects and would consequently be predicted to have derived from objects with initial masses of 3$-$4 M$_{\sun}$.  
This is close to the upper limit that one would expect for a carbon-rich star. 

\subsection{Post-AGB Objects (PPNe))}

Periodic light variations have been determined for the first time for two of the PPNe, with periods of 96 and 123 days.
These are listed in Table~\ref{results}, which also includes the brightness, maximum seasonal ranges in brightness, spectral types, periods, period ratios, 
and some comments for each of the program objects. 
Previous studies have shown that PPNe typically range in periodic variability from as short as $\sim$35 days for early-F spectral types to as long as 160 days for late-G spectral types, with early-B types showing variability on the order of days without any reported periodicity \citep{hri15b,ark13}.
The two PPNe with periods have spectral types from F6 and G2, and their periods fit nicely within that period range.
The other PPN, IRAS 19306+1407, has an early B spectral type and shows variation, peak-to-peak, on the order of several days, consistent with what is seen in other hot PPNe.

\placetable{results}
 
\begin{figure}\figurenum{8}\epsscale{1.00} 
\plotone{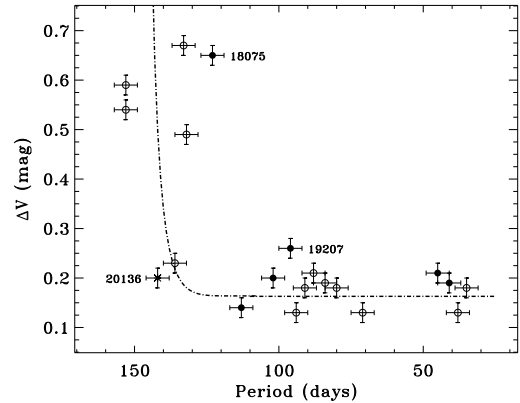}
\caption{The relationship between period and maximum seasonal variation ($\Delta${\it V}) for our two new PPNe, compared with previously-studied PPNe.  O-rich objects are shown with filled circles and C-rich with open circles.  We have assumed uncertainties of $\pm$3 days (conservative) and $\pm$0.02 mag.   The dashed line is simply a free-hand representation of the trend in the data.  Also included is IRAS 20136+1309 (x), which appears to be a post-RGB object.
\label{P-dV}}
\epsscale{1.0}
\end{figure}

Unfortunately, none of these objects has an accurate photospheric temperature determined by analysis of high-resolution spectra.
Thus we cannot explicitly compare them to the {\it T}$_{eff}$$-$P relationship found for PPNe.
They do agree with the general trend of cooler systems having longer periods.
We can, however, compare them to the relationship found between period and amplitude for PPNe \citep{hri15b}.  This is shown in Figure~\ref{P-dV}, where for amplitude we have plotted the maximum variation $\Delta${\it V} in a season.  
Neither IRAS 18075$-$0924 or 19207+2023 fit the trend very well, but they do show the same general characteristics, with the larger amplitude variation associated with the longer period.
Note that the maximum variation in brightness seen in IRAS 18075$-$0924 is much larger than that documented for any of the previously analyzed O-rich sources. 

\placefigure{P-dV}

Previous studies of PPNe have demonstrated that these variations are due to pulsations \citep{hri13,hri18}.  It has been found that the brightness and color reached minimum values at nearly the same phases, with the objects faintest when reddest, while the maximum in the radial velocity differed by a quarter of a cycle \citep{hri13,hri18}.  
For the objects in the present study, we don't have radial velocity curves nor do we have color curves that show periodicity.   However, we did compare the {\it V} and {\it R}$_C$ light curves, and found, when adopting the ephemeris of the {\it V} light curve, that the {\it V} and {\it R$_C$} light curves reached minimum light at the same phase, to within the phase uncertainties ($\pm$0.02$-$0.03).  This is consistent with the phase relationship found in previous observational studies of PPNe.

Previous studies of PPNe also found multiple periods in the light curve variations, commonly with the second period close to the first.  Typical period ratios, {\it P}$_2$/{\it P}$_1$, ranged from 0.8 to 1.2 \citep{hri13,hri15b}.  IRAS 19207+2023 has a similar period ratio of 1.05.
IRAS 18075$-$0924 has a secondary period of 347 days, suspiciously close to our one year observing cycle. 
If we ignore this and use instead the third period, we find a ratio of 0.87.  
Thus, with the caveat described above, the period ratios for these two PPNe are in agreement with those found previously. 

Two of the objects show evidence of multi-year periodicity.  
IRAS 18075$-$0924 displays a period of 3.5 yrs in both the {\it V} and {\it R}$_C$ light curves, which agrees with the timing of the three deeper minima.  IRAS 19207+2023 has a period of 4.2 yrs in the {\it V} light curve, which can be seen as a more gradual wave through the light curve, peaking in 2011 and 2015. 
One might speculate that these longer-term periodicities are due to the oblate shape of the stars caused by the tidal forces of an unseen companion, in which case the orbital period would be twice these values.  However, for stars with these luminosities and temperatures, the radii of the stars ($\sim$60-100 R$_\sun$) are too small to produce even the small semi-amplitude (0.027 mag) seen in IRAS 19207+2023.
A better explanation for these multi-year periodicities might involve an unseen binary companion, but with the periodic light variations due to obscuration as the PPNe orbits within a circumbinary disk. 
This would be interesting, as at present, there are no well-supported cases of PPNe with binary companions.
There has been a long-term radial velocity program to monitor a sample of the brightest of these but with no clear detections \citep{hri11,hri17}.
These light variations could also be due to variations in the circumstellar extinction with time.
Continued monitoring of these two stars for another decade, perhaps with the ASAS-SN or some similar survey, would be very helpful to verify or refine these periods.

\subsection{Post-RGB Objects}

Two of the objects, IRAS 19039+1232 and 20136+1309, are candidates for objects in the post-RGB phase.
This is based on their low luminosity derived from {\it Gaia} (DR2) distances, even with the distance uncertainties included.
One might argue that the spectroscopicly-assigned luminosity classifications are in disagreement with this, since both are classified as I, supergiant, rather than III, giant.  However, \citet{kam14,kam15,kam16} found that most of the post-RGB objects in the Magellanic Clouds have log~{\it g} values of 0 to 2, consistent with a supergiant classification.  
These two objects are large, with luminosities of approximately 1000$-$1500 L$_{\sun}$, and they have also lost considerable mass, as attested by their large IR excesses.  Together these would contribute to a small value of log~{\it g}.

Light variations are seen in each of them. 
For IRAS 20136+1309, with mid-F spectral type, a dominant period of 142 days has been found, with a period ratio of 0.95.
IRAS 19039+1232, with a mid-B spectral type, shows variations, peak-to-peak, over the course of a few days.
These results have also been included in Table~\ref{results}.

Long-term, approximately monotonic trends in brightness are observed in both objects.  
IRAS 19039+1232 decreased in brightness from 2008 to 2017 by 0.06 mag or 0.006 mag yr$^{-1}$ and  
IRAS 20136+1309 increased by 0.11 mag or 0.012 mag yr$^{-1}$ over this same interval.
The ({\it V$-$R}$_C$) color of IRAS 19039+1232 did not change, while IRAS 20136+1309 got redder by about 0.02 mag as it brightened.  
Suggested causes of the brightness changes are the evolution of the star or changes in the effective extinction from the circumstellar envelope.  
The fact that IRAS 20136+1309 got redder as it brightened is not easily explained by either of these causes.
Nor is there evidence of H$\alpha$ emission from this object that could be increasing as it brightens, causing the {\it R}$_C$ band brightness to increase.  

Dusty post-RGB objects are expected to have evolved from binaries, in which the companions interrupted their evolution.
However, in contrast to two of the other program objects which appear to be PPNe, neither of these post-RGB candidates shows long-term variations in its light curve that might be an indication of a binary nature.
The SED of IRAS 20136+1309 shows excess emission in the near-infrared, evidence of warm dust, perhaps at a range of temperatures., in addition to the cooler dust emitting in the mid-infrared.  This is suggestive of a disk around the star. 
Radial velocity studies are likely need to verify any binary companion.

\section{RESULTS AND SUMMARY}
\label{summary}

We have carried out a photometric study of five evolved stars of intermediate-brightness ({\it V} $\approx$ 13$-$15 mag) which possesses large mid-infrared excesses.  Observations were carried out over a 10 or 11 year time interval using three different telescopes.  Recent observations covering three or four years from the ASAS-SN survey were also examined.   
All five vary in brightness, three of them periodically.  
Chemically, two of the objects are clearly classified as O-rich, one as likely O-rich, one as mixed O-rich and C-rich, and one as uncertain.
Below are listed the primary results of this study. 

1. Three of these objects vary periodically in light, with dominant periods ranging from 96 to 142 days.  The other two are hotter and have variations on the time scale of a few days. 
The maximum seasonal variations are small for four of them, only 0.13$-$0.26 mag in {\it V} and less in {\it R}$_C$.  One, however, is much larger, with a maximum of 0.65 mag ({\it V}). 

2. Luminosities were calculated based on Gaia distances, and for two of the objects the luminosities were low, 1000$-$1500 L$_\sun$.  This led to a classification of the two as post-RGB rather than post-AGB PPNe.  The other three fit the luminosity range of PPNe.

3. The two PPNe with periods fit the general trend for PPNe of having shorter periods and smaller amplitudes with earlier spectral types (higher temperatures).

4. Both of the periodic PPNe also possess secondary periods not differing much from the primary periods, with period ratios ({\it P}$_2$/{\it P}$_1$) = 0.89$-$1.06.  
These are similar to the values found previously for other PPNe.

5. Two of the PPNe give some evidence of longer periodicities: IRAS 18075$-$0924 (3.5 years) and IRAS 19207+2023 (4.2 years).  
Ongoing observations, such as those from the ASAS-SN survey, will be useful to investigate if these are indeed periodic or not.  If indeed periodic, then they might be photometric evidence of binarity in these PPNe.

6. The two post-AGB candidates display approximately monotonic changes in brightness: IRAS 20136+1309 increased in brightness by 0.11 mag and IRAS 19039+1232 decreased by 0.06 mag over the 10 years of observations.  However, neither of these show any long-term periodicity suggestive of binarity in their light curves.
 
 As a result of this study, we have added information on three additional PPNe to our growing knowledge of the pulsational properties of this subclass of post-AGB stars.
The other two stars, though similar in many ways, appear to have much lower luminosities, and thus fit better into the recently identified class of dusty post-RGB objects \citep{kam16}.  These two are among the first to be identified in the MWG.  As such, they, would repay further study, in particular high-resolution spectroscopy to study their atmospheric properties.  Radial velocity monitoring of IRAS 20136+1309, with its mid-F spectral type and low interstellar extinction, would appear to make it a favorable target for a search for binarity.  

\acknowledgments

We gratefully acknowledge the many Valparaiso University undergraduate students who carried out the photometric observations at the VU Observatory from 2008 to 2018 that were used in this study.
They are, in chronological order, Ryan McGuire, Sam Schaub, Chris Wagner, 
Zach Nault, Wesley Cheek, Joel Rogers, Rachael Jensema, Chris Miko, Austin Bain, Hannah Rotter, Aaron Seider, Allyse Appel, Brendan Ferris, Justin Reed, Jacob Bowman, Ryan Braun, Dani Crispo, Stephen Freund,  Chris Morrissey, Cole Hancock, Abigail Vance, Kathryn Willenbrink, Andrew Webb, Matthew Bremer, Avery Jackson, David Vogl, Tim Bimler, and Sammantha Nowak-Wolff.
We also thank R. Kaitchuck and B. Murphy for obtaining some SARA observations used in this study and VU undergraduates Justin Reed, Matthew Bremer, and David Vogl for assistance in the reduction of the various photometry sets. 
We thank the referee for insightful suggestions and a helpful reference that improved our study.
The ongoing work of Paul Nord in maintaining the VU Observatory and assisting in software is gratefully acknowledged.
BJH acknowledges ongoing support from the National Science
Foundation (0407087, 1009974, 1413660) and the Indiana Space Grant Consortium.
This research has made use of the SIMBAD and VizieR databases, operated at CDS, Strasbourg,
France, and NASA's Astrophysical Data System.

\section{Appendix: Combining the Multi-telescope Observations}

The use of multiple telescopes at different latitudes and longitudes has allowed us to obtain more data than could be done using only the VUO telescope.  
It has, however, introduced some complications.
When we initially combined the different data sets for the particular objects, some systematic offsets were apparent in the light curves.  
These are thought to arise from the neglect in the standardization of second-order effects in the color and the extinction terms for these very red program objects, and in some cases  
 a relatively large color coefficient (0.1$-$0.2) for a detector-filter system.
Since all of the objects were observed over the entire 11-year observing interval at the VUO using the same detector and filter set, we adopted the VUO data as the reference system.
We then carefully determined empirical offset values for the other telescope-filter-detector systems.  This was done primarily by determining the differences in the magnitudes of observations made at both VUO and the other telescope-filter-detector systems on the same or adjacent nights.
Since the objects are typically seen to vary by small amounts over intervals of weeks or months, we expect them to remain essentially unchanged in brightness over this one day interval.  
While this is less the case for IRAS 19039+1232 and 19306+1407, which vary on shorter timescales, it is the best that we could do.
We note that most of the offsets for these two objects are small ($\le$0.03 mag in absolute value).
In a few cases, when such observations were not available, we boot-strapped by comparing one data set with another one for which the offset had been determined.  
These systematic offset values were generally small, between $-$0.02 and $+$0.02 mag, but in a few cases they reached to $\pm$0.10 mag. and in one case to $-$0.14 mag.
These were all checked visually to confirm that the adopted offsets did indeed improve the compatibility of the different data sets.
These larger ones are clearly needed to remove obviously large differences from the VUO dataset. 
The offset values used to bring the various SARA telescope-detector-filter data to the VUO levels for each star are listed in Table~\ref{offsets}.
Details of the SARA telescopes and detectors are given by \citet{keel17}.
We estimate the uncertainty in these offset values to be $\pm$0.01 to $\pm$0.02 mag.  

\placetable{offsets} 

The photometric data for the five program stars, with the offsets included, are listed in Table~\ref{std_mags}, which is available in its entirety in machine-readable form.  Included are the heliocentric Julian date (HJD) of the observation, the standardized differential magnitude (program star $-$ comparison star 1), and a code to identify the particular telescope-detector-filter set used.
The codes are defined in Table~\ref{offsets}.
We have included the earlier 1994$-$2007 observations of IRAS 20136+1309.

\placetable{std_mags}

\clearpage

\tablenum{1}
\begin{deluxetable}{clrrrccccl}
\rotate
\tablecaption{List of PPN Targets Observed \label{object_list}}
\tabletypesize{\footnotesize} \tablewidth{0pt} \tablehead{
\colhead{IRAS ID}&\colhead{GSC ID\tablenotemark{a}}&\colhead{2MASS ID}&\colhead{R.A.\tablenotemark{b}}&\colhead{Decl.\tablenotemark{b}}
&\colhead{{\it l}}&\colhead{{\it b}}
&\colhead{{\it V}\tablenotemark{c}}&\colhead{{\it B$-$V}\tablenotemark{c}}&\colhead{Sp.T.} \\
&&&(2000.0)&(2000.0)&($\arcdeg$)&($\arcdeg$)&\colhead{(mag)} &\colhead{(mag)} & } 
\startdata
18075$-$0924 & S9NM000112 & J18101514$-$0923350   & 18:10:15.14 & $-$09:23:35.1 & 019.8 & +04.7    & 13.9  & 2.5 & G2~I\tablenotemark{d} \\
19039+1232   & N2BR033916    & J19061532+1236488     & 19:06:15.32 & +12:36:48.8   & 045.8 & +02.5    & 14.7  & 1.7 & B5:~Ie\tablenotemark{e} \\
19207+2023   & N27X026988    & J19225582+2028547     & 19:22:55.82 & +20:28:54.7   & 054.7 & +02.6    & 15.2   & 2.4: & F6~I\tablenotemark{d} \\
19306+1407   & N277049412    & J19325508+1413369     & 19:32:55.08 & +14:13:37.0   & 050.3 & $-$02.5 & 14.2  & 1.3 & B0-1~Ie\tablenotemark{f} \\
20136+1309   & N1UR032398    & J20160051+1318562     & 20:16:00.51 & +13:18:56.3   & 054.8 & $-$12.0  & 13.4 & 1.0 & F3-7~I\tablenotemark{f}, F6~I\tablenotemark{e} \\
\enddata
\tablenotetext{a}{{\it Hubble Space Telescope} Guide Star Catalog II (GSC-2), version 2.3.2 (2006).}
\tablenotetext{b}{Coordinates from the 2MASS Catalog.}
\tablenotetext{c}{These values are all variable as discussed in this paper.  Uncertain values are listed with colons.}
\tablenotetext{d}{Spectral types by \citet{suarez06}.  However, we regard their classification of IRAS 19306+1407 as G5~I to be in error; see discussion by \citet{sancon08}. } 
\tablenotetext{e}{Based on our unpublished spectra; note that we have not included the uncertain spectral types listed with colons by \citet{kel05}. }
\tablenotetext{f}{Spectral types by \citet{sancon08}. }
\end{deluxetable}


\tablenum{2}
\begin{deluxetable}{lrrrrl}
\tablecaption{Observed Standard Magnitudes and Colors of the Program Stars
\label{std_ppn}}
\tabletypesize{\footnotesize} 
\tablewidth{0pt} \tablehead{ \colhead{IRAS ID} &\colhead{{\it V}}
 &\colhead{{\it B$-$V}} &\colhead{{\it V$-$R$_C$}} &\colhead{{\it V$-$I$_C$}} 
&\colhead{Date}  \\
&\colhead{(mag)} & \colhead{(mag)} & \colhead{(mag)} & \colhead{(mag)} 
&\colhead{} } 
\startdata
18075$-$0924 	& 14.03  & 2.54 & 1.54 & 2.89 &  1995 Sep 13\tablenotemark{a}   \\
		 	& 13.72  & \nodata & 1.45 & \nodata &  2009 May 20  \\
		        & 13.91   & \nodata &  1.54 & \nodata & 2013 Jul 17  \\              
                         & 13.81   & \nodata &  1.52 & \nodata & 2013 Aug 21  \\ 
                         & 14.26  & \nodata & 1.51 & \nodata &  2014 Jun 06  \\
19039+1232\tablenotemark{b}   & 14.70   & 1.68 & 1.13 & 2.28 & 1994 Jun 23\tablenotemark{a}  \\ 
			& 14.66   & \nodata & 1.05 & \nodata &  2009 May 20 \\ 
                        & 14.76   & \nodata &  1.05 & \nodata & 2009 Nov 11  \\ 
                         & 14.75  & \nodata & 1.09 & \nodata & 2009 Nov 25  \\
19207+2023    & 15.36:  & 2.4: & 1.54:  & 2.97: & 1995 Sep 13  \\
			& 15.16  & \nodata & 1.42  & \nodata & 2009 May 20  \\
                         & 15.26 & \nodata & 1.49 & \nodata & 2009 Nov 11 \\
                         & 15.17 & \nodata & 1.49 & \nodata & 2009 Nov 25 \\
                         & 15.26 & \nodata & 1.50  & \nodata & 2014 Jun 06  \\
19306+1407\tablenotemark{b}  & 14.19 & 1.26 & 0.83 & 1.68 & 1994 Jun 23\tablenotemark{a}  \\ 
			& 14.10 & \nodata & 0.82 & \nodata & 2009 May 20 \\ 
                         & 14.14 & \nodata & 0.83 & \nodata & 2009 Nov 11 \\
                        & 14.22 & \nodata & 0.84 & \nodata & 2009 Nov 25 \\
                         & 14.19  & \nodata & 0.83 & \nodata & 2014 Jun 06  \\
20136+1309  & 13.31 & 0.96 & \nodata  & 1.44 & 1989 Oct 01\tablenotemark{c} \\ 
		     & 13.40 & \nodata & 0.73  & \nodata & 2009 May 20 \\ 
\enddata
\tablecomments{Uncertainties in the brightness and color are $\pm$0.015$-$0.02 mag, except for those indicated with colons, which have larger uncertainties.}
\tablenotetext{a}{Observed earlier at Kitt Peak National Observatory.}
\tablenotetext{b}{Note that the {\it R}$_C$ measurement contains a contribution from H$\alpha$ emission.}
\tablenotetext{c}{~Listed by \citet{su01}.}
\end{deluxetable}


\tablenum{3}
\begin{deluxetable}{llrrrr}
\tablecaption{Comparison Star Identifications and Standard Magnitudes \label{std_comp}}
\tabletypesize{\footnotesize}
\tablewidth{0pt} \tablehead{\colhead{IRAS Field}
&\colhead{Object} &\colhead{GSC ID} &\colhead{2MASS ID} &\colhead{{\it V}} 
&\colhead{{\it V$-$R}$_C$}  \\
\colhead{}
&\colhead{} &\colhead{} &\colhead{} &\colhead{(mag)} 
 &\colhead{}} \startdata
18075$-$0924 & C1 & S9NM000118 & 18100090$-$0926081 & 12.58 & 0.55  \\
              & C2 & S9NM000114 & 18102436$-$0923491 & 12.48 & 0.57  \\
              & C3 & S9NM000077 & 18103008$-$0922288 & 13.95& 1.42   \\
19039+1232  & C1 & N2BR035531 & 19060538$+$1237588 & 14.46 & 0.37 \\
              & C2 & N2BR031476 & 19060419$+$1235036 & 14.21 & 0.53   \\
              & C3 & N2BO081294 & 19062533$+$1237143 & 14.77 & 0.84  \\
19207+2023   & C1 & N27X028074 & 19225394$+$2030146 & 14.82  & 0.74 \\
              & C2 & N27X027650 & 19224995$+$2029440 & 15.18  & 1.21  \\
              & C3\tablenotemark{a} & N27X024824 & 19225290$+$2026404 & 14.33  & .0.62 \\
19306+1407  & C1 & N277050167 & 19331311$+$1414432 & 13.58 & 0.57 \\
              & C2 & N276000327 & 19324127$+$1416568 & 13.52 & 0.98 \\
              & C3 & N276000333& 19324186$+$1413158 & 12.64 & 0.30 \\
20136+1309  & C1 & N1UR032383 & 20160534$+$1318516 & 13.25 & 0.34 \\
              & C2 & N1UR031931 & 20160537$+$1318223 & 13.15 & 0.38 \\
              & C3 & N1UR031731 & 20160317$+$1318049 & 13.28 & 0.67  \\
\enddata
\tablecomments{Uncertainties in the brightness and color are $\pm$0.01$-$0.02 mag.}
\tablenotetext{a}{The brightness of C$_3$ decreases approximately monotonically by 0.02 mag ({\it R}$_C$) from 2008 to 2017.}
\end{deluxetable}


\tablenum{4}
\begin{deluxetable}{lcrrrrrrrrrrrrrrr}
\tablecolumns{17} \tabletypesize{\scriptsize}
\tablecaption{Periodogram Study of the Light and Color Curves\tablenotemark{a,b}\label{periods}}
\rotate
\tabletypesize{\footnotesize} 
\tablewidth{0pt} \tablehead{ 
\colhead{IRAS ID} &\colhead{Filter} & \colhead{Years}&\colhead{No.} & \colhead{{\it P}$_1$}&\colhead{{\it A}$_1$} &\colhead{{\it $\phi$}$_1$\tablenotemark{c}}&\colhead{{\it P}$_2$}&\colhead{{\it A}$_2$} &\colhead{$\phi$$_2$\tablenotemark{c}}
&\colhead{{\it P}$_3$} &\colhead{{\it A}$_3$}&\colhead{$\phi$$_3$\tablenotemark{c}}&\colhead{{\it P}$_4$} &\colhead{{\it A}$_4$}&\colhead{$\phi$$_4$\tablenotemark{c}}&\colhead{$\sigma$\tablenotemark{d}}\\
 & & &\colhead{Obs.} & \colhead{(days)}&\colhead{(mag)} & &\colhead{(days)}&\colhead{(mag)}
 & &\colhead{(days)} &\colhead{(mag)}& &\colhead{(days)} &\colhead{(mag)} & &\colhead{(mag)}}
\startdata
18075$-$0924 & {\it V}\tablenotemark{e} & 2009-2018 & 171 & 123.1 & 0.076 & 0.40 & 346.2 & 0.095 & 0.73 & 107.7 & 0.062 & 0.78 & 209.1 & 0.051 & 0.22 & 0.065 \\
18075$-$0924 & {\it R$_C$}\tablenotemark{e} & 2009-2018 & 192 & 122.9 & 0.081 & 0.40 & 346.9 & 0.113 & 0.74 & 107.7 & 0.063 & 0.77 & 208.8 & 0.049 & 0.22 & 0.065 \\
\\
19207$+$2023  & {\it V}\tablenotemark{e} & 2008-2017 & 229 & 95.9 & 0.046 & 0.10 & 93.2: & 0.022 & 0.64 & \nodata & \nodata & \nodata & \nodata & \nodata & \nodata & 0.042 \\
19207$+$2023  & {\it R$_C$}\tablenotemark{e} & 2008-2017 & 261 & 95.7 & 0.033 & 0.06 & 100.3 & 0.021 & 0.07 & 114.4 & 0.021 & 0.39 & 85.0 & 0.017 & 0.02 & 0.030 \\
\\
20136$+$1309 & {\it V}\tablenotemark{e,f} & 2008-2017 & 140 & 142.3 & 0.026 & 0.56 & 135.7& 0.021 & 0.91 & 68.8 & 0.020 & 0.34  & \nodata & \nodata & \nodata & 0.027 \\
20136$+$1309 & {\it R$_C$}\tablenotemark{e,f} & 2008-2017 & 148 & 135.1 & 0.020 & 0.89 & 142.8 & 0.020 & 0.57 & 70.9 & 0.017 & 0.23 & \nodata & \nodata & \nodata & 0.024 \\
20136$+$1309 & {\it V}\tablenotemark{e} & 1994-2007 & 104 & 126.7 & 0.049 & 0.55 &138.0 & 0.034 & 0.11 & \nodata & \nodata & \nodata  & \nodata & \nodata & \nodata & 0.039 \\
20136$+$1309 & {\it V}\tablenotemark{e,f} & 1994-2017 & 236 & 143.2 & 0.026 & 0.62 & 127.0 & 0.028 & 0.47 & 135.1& 0.021 & 0.90  & \nodata & \nodata & \nodata & 0.035 \\
20136$+$1309 & {\it V}\tablenotemark{e,g} & 2015-2017 & 170 & 146.0 & 0.046 & 0.88 & 320.0 & 0.033 & 0.10 & \nodata & \nodata & \nodata & \nodata & \nodata & \nodata & 0.026 \\
\enddata
\tablenotetext{a}{The uncertainties in {\it P}, {\it A}, $\phi$ are $\pm$0.1$-$0.6 days, $\pm$0.003$-$0.007 mag, $\pm$0.01$-$0.03, respectively.  Exceptions are the larger uncertainty ($\pm$0.8 days) in {\it P}$_1$ found in the shorter time interval ASAS-SN light curve of IRAS 20136+1309 and the larger uncertainties ($\pm$1$-$2 days) found in the longer periods: {\it P}$_2$ and {\it P}$_4$ in IRAS 18075$-$0924 and {\it P}$_2$ in the ASAS-SN light curve of IRAS 20136+1309.}
\tablenotetext{b}{Colons (:) indicate less certain period values that fell slightly below our adopted level of significance; see text for details.}
\tablenotetext{c}{The phases are determined based on the epoch of 2,455,600.0000, and they each represent the phase derived from a sine-curve fit to the data, not the phase of minimum light.}
\tablenotetext{d}{Standard deviation of the observations from the sine-curve fit.}
\tablenotetext{e}{Analysis based on the seasonally normalized light curve.}
\tablenotetext{f}{Analysis based on the light curve with the long-term trend first removed.}
\tablenotetext{g}{ASAS-SN data.}
\end{deluxetable}


\tablenum{5}
\begin{deluxetable}{rrrrcrrcl}
\tablecaption{Calculated Luminosities of the Program and Related Objects
 \label{sed_table}}
\tabletypesize{\footnotesize}
\tablewidth{0pt} \tablehead{ \colhead{IRAS ID} & \colhead{{\it Gaia} DR2 ID} &\colhead{{\it D}\tablenotemark{a}} &\colhead{{\it A}$_V$(C)\tablenotemark{b}} 
&\colhead{{\it A}$_V$(SF)}  &\colhead{{\it L}(C)\tablenotemark{c}} &\colhead{{\it L}(SF)\tablenotemark{c}} & Chem & \colhead{Type} \\
\colhead{} &\colhead{} &\colhead{(kpc)} &\colhead{(mag)}&\colhead{(mag)} &\colhead{(L$_{\sun}$)} &\colhead{(L$_{\sun}$)} &\colhead{} &\colhead{} }
\startdata
 \multicolumn{9}{c}{Program Objects} \\
  \tableline 
18075$-$0924 & 4158154754919296000 & 5.70 & 3.47 & 3.74 & 6400 & 7000 & O & post-AGB\\ 
19039+1232    & 4313970228532767360 & 3.44 & 2.3: & 4.21 & 800: & 1100 & O & post-RGB \\
19207+2023    & 4516723883521069952 & 5.81 & 2.17 & 5.62 &2600 & 5800 & O & post-AGB \\
19306+1407    & 4318134628803970816 & 5.58 & 2.17 & 4.52  & 9500 & 12,000 & C-O & post-AGB \\
20136+1309    & 1803364717856260736 & 4.09 & 0.37 & 0.53 & 1600 & 1600 & O & post-RGB\\
 \tableline 
 \multicolumn{9}{c}{Similar Objects} \\
 \tableline 
04296+3429    & 173086700992466688 & 2.98 & 2.2: & 2.06  & 3100: & 3000 & C & post-AGB\\ 
06530$-$0213  & 3105987960396950784 & 4.01 & 0.96 & 2.96 & 3500 & 3700 & C & post-AGB \\
23304+6147    & 2015785313459952128 & 7.17 & 3.86 & 4.18 & 27,000 & 29,000 & C & post-AGB \\
17436+5003    & 1367102315248484864 & 3.16 & 0.09 & 0.08 & 20,000 & 20,000 & O & post-AGB \\
18095+2704    & 4580154606223711872 & 3.64 & 0.23: & 0.27  & 14,000: & 14,000 & O & post-AGB\\
19386+0155    & 4240112390324832384 & 4,64 & 0.99 & 1.08  & 16,000 & 16,000 & O & post-AGB \\
19475+3119    & 2033763428091006720 & 2.84 & 1.46 & 3.13  & 4800 & 13,000 & O & post-AGB \\
\enddata
\tablecomments{{\it A}$_V$(C) and {\it A}$_V$(SF) refer to the {\it V} extinction as determined from \citet{cap17} and \citet{sf11}, respectively.  
{\it L}(C) and {\it L}(SF) refer to the luminosity determined with the \citet{cap17} value and with the \citet{sf11} value, respectively.
Note that these values of {\it A}$_V$ are in general agreement with those from \citet{vic15} except for IRAS 19306+1407, where they get an especially low value for an object in the galactic plane.}
\tablenotetext{a}{Uncertainties in the distances are large; they typically have a lower limit of $\sim$0.7 and an upper limit of $\sim$1.5 times the distance listed.  Somewhat smaller ranges in uncertainties of $\sim$0.85 and $\sim$1.2 times are found for IRAS 06530$-$0213, 17436+5003, and 19475+3119.}
\tablenotetext{b}{The colons indicate more uncertain values.}
\tablenotetext{c}{The uncertainties in the luminosities, based on the uncertainties in the distances, range from a lower limit of a factor of $\sim$0.45 to an upper limit of a factor of $\sim$2.30 for each of the objects, except for the three with smaller distance uncertainties.  For IRAS 06530$-$0213, 17436+5003, and 19475+3119, the ranges are smaller, $\sim$0.7 and $\sim$1.5.}
\end{deluxetable}


\tablenum{6}
\begin{deluxetable}{rrcrrrrl}
\tablecaption{Results of the Period and Light Curve Study 
 \label{results}}
\tabletypesize{\footnotesize}
\tablewidth{0pt} \tablehead{ \colhead{IRAS ID} &\colhead{V} &\colhead{$\Delta$V\tablenotemark{a}} 
&\colhead{SpT} &\colhead{P$_1$} &\colhead{P$_2$} &\colhead{P$_2$/P$_1$} & \colhead{Comments} \\
\colhead{} &\colhead{(mag)} &\colhead{(mag)}&\colhead{} &\colhead{(days)} &\colhead{(days)} &\colhead{}  &\colhead{} }
\startdata
 \multicolumn{8}{c}{PPNe} \\
\tableline 
18075$-$0924 & 14.0 & 0.65 & G2~I & 123 & 346 & 0.87\tablenotemark{b} & P$_3$=108 days\\
19207+2023   & 15.2 & 0.26 & F6~I & 96 &100 & 1.05  & \nodata \\
19306+1407   & 14.2 & 0.13 & B0-1~Ie & \nodata & \nodata & \nodata & \nodata \\
\tableline 
\multicolumn{8}{c}{post-RGB} \\
\tableline 
19039+1232   & 14.7 & 0.16 & B5:~Ie & \nodata & \nodata & \nodata  & Trend of decreasing brightness \\
20136+1309   & 13.2 & 0.20 & F3-7~I & 142 & 136 & 0.95\tablenotemark{c} & Trend of increasing brightness\\
\enddata
\tablenotetext{a}{The maximum brightness range observed in a season.}
\tablenotetext{b}{P$_3$/P$_1$.}
\tablenotetext{c}{Value in {\it V} light curve; P$_2$/P$_1$=1.06 in {\it R}$_C$ light curve.  Using the full 1994$-$2017 {\it V} light curve results in P$_2$/P$_1$=0.89}
\end{deluxetable}


\tablenum{7}
\begin{deluxetable}{lclllllllllllllll}
\rotate
\tablecaption{SARA Telescope-Detector-Filter Offsets for Each Star\tablenotemark{a}\label{offsets}}
\tabletypesize{\footnotesize} \tablewidth{0pt} \tablehead{
\colhead{Telescope-}&\colhead{Code\tablenotemark{b}}&&\multicolumn{2}{c}{IRAS 18075}&&\multicolumn{2}{c}{IRAS 19039}&&\multicolumn{2}{c}{IRAS 19207}&&\multicolumn{2}{c}{IRAS 19306}  &&\multicolumn{2}{c}{IRAS 20136} \\
\cline{4-5} \cline{7-8}  \cline{10-11} \cline{13-14}  \cline{16-17}
\colhead{Detector}&\colhead{}&&\colhead{{\it V}}&\colhead{{\it R}$_C$}&&\colhead{{\it V}}&\colhead{{\it R}$_C$}&&\colhead{{\it V}}&\colhead{{\it R}$_C$}
&&\colhead{{\it V}}&\colhead{{\it R}$_C$}&&\colhead{{\it V}}&\colhead{{\it R}$_C$}\\
\colhead{}&\colhead{}&&\colhead{(mag)}&\colhead{(mag)}&&\colhead{(mag)}&\colhead{(mag)}
&&\colhead{(mag)}&\colhead{(mag)}&&\colhead{(mag)}&\colhead{(mag)}&&\colhead{(mag)}&\colhead{(mag)}}
\startdata
 VUO 	SBIG & A && $+$0.00 & $+$0.00  && $+$0.00 & $+$0.00 && $+$0.00 & $+$0.00  && $+$0.00 & $+$0.00 && $+$0.00 & $+$0.00 \\
SARA-KP U42   & B && $-$0.01 & $+$0.005 && $-$0.02 & $+$0.01 && $-$0.01 & $+$0.01  && $+$0.01 & $+$0.03 && \nodata   & \nodata\\
SARA-KP U42\tablenotemark{c} & C && $-$0.06: & $+$0.02:   && \nodata  & \nodata  &&  \nodata & +0.00  && \nodata & \nodata && \nodata   & \nodata\\
SARA-KP FLI\tablenotemark{d}  & D && \nodata  & \nodata   && $-$0.04 & $-$0.04&& \nodata  & \nodata  && \nodata   & \nodata && \nodata   & \nodata\\
SARA-KP ARC  & E && $+$0.01 & $+$0.055 && $-$0.015 & $+$0.06 && $-$0.03 & $+$0.03  && $+$0.01 & $+$0.07 && \nodata   & \nodata\\
SARA-CT E6     & F && $-$0.14 & $-$0.10  && \nodata   & \nodata &&  $-$0.09: & $-$0.06  &&  \nodata  & \nodata && \nodata   & \nodata\\
SARA-CT ARC & G && $-$0.02 & $+$0.00 && \nodata  & \nodata &&  \nodata   & \nodata   &&  \nodata  & \nodata && \nodata   & \nodata\\
SARA-CT FLI   & H && $-$0.085 & $+$0.01 && \nodata  & \nodata &&  \nodata   & \nodata   &&  \nodata  & \nodata && \nodata   & \nodata\\
VUO PHOT	& I && \nodata  & \nodata   && \nodata  & \nodata && \nodata  & \nodata  && \nodata   & \nodata &&  $+$0.00 & $+$0.00 \\
\enddata
\tablecomments{Uncertainties are $\pm$0.01 to $\pm$0.02 mag; those with colons (:) are the more uncertain.  No entry in a column indicates that no observations were made or used for that star with that telescope-detector-filter set.}
\tablenotetext{a}{Offsets as compared to the VUO values, in the sense offset value = mag(VUO) $-$ mag(system).  
Thus these offset values were added to the SARA magnitudes to bring them to the VUO SBIG photometric system. }
\tablenotetext{b}{Represents the coding used in the Table~\ref{std_mags} to identify the source of the photometry and the associated offset for each star. }
\tablenotetext{c}{Alternate filter set. }
\tablenotetext{d}{Finger Lakes camera used at SARA-KP for a short time in 2009. }
\end{deluxetable}


\tablenum{8}
\begin{deluxetable}{llclcc}
\tablecaption{Differential Standard Magnitudes \label{std_mags}}
\tablewidth{0pt} \tablehead{\colhead{IRAS ID}
&\colhead{HJD({\it V})} &\colhead{$\Delta${\it V}} &\colhead{HJD({\it R}$_C$)} &\colhead{$\Delta${\it R}$_C$} 
&\colhead{Code\tablenotemark{a}} \\
\colhead{} &\colhead{} &\colhead{(mag)} &\colhead{} 
&\colhead{(mag)} & \colhead{}} 
\startdata
IRAS18075-0924 &                     & 	       & 54665.6538 & 0.422 &  A \\
IRAS18075-0924 & 54671.6553 & 1.440 & 54671.6416 & 0.479 &  A \\
IRAS18075-0924 & 54676.6567 & 1.429 & 54676.6434 & 0.476 &  A \\
IRAS18075-0924 & 54690.6071 & 1.365 & 54690.5938 & 0.414 &  A \\
IRAS18075-0924 & 54693.6806 & 1.314 & 54693.6650 & 0.370 &  A \\
IRAS18075-0924 & 54694.6694 & 1.315 & 54694.6541 & 0.380 &  A \\
IRAS18075-0924 & 54698.5902 & 1.313 & 54698.5756 & 0.388 &  A \\
IRAS18075-0924 & 54719.6125 & 1.249 & 54719.5997 & 0.300 &  A \\
IRAS18075-0924 &                     &   	      & 54744.5423 & 0.354 & A \\
IRAS18075-0924 & 54969.8710 & 1.231 & 54969.8586 & 0.273 & A \\
\enddata
\tablecomments{Typical uncertainty in the standardized differential magnitudes is $\pm$0.015 mag.}
\tablenotetext{a}{This identifies the source of the photometry, as listed in Table~\ref{offsets}, and the associated added offset, if any, to bring that observation to the VUO system. \\ (This table is available in its entirety in machine-readable form.)}
\end{deluxetable}

\clearpage

\clearpage

\clearpage

\clearpage

\clearpage

\clearpage

\end{document}